\documentclass[sigconf]{acmart}

\usepackage{microtype}
\usepackage{graphicx}
\usepackage{booktabs} 
 \usepackage{tikz}
 \usepackage{pifont}
\usepackage{amsmath}
 
\usepackage{mathtools}
\usepackage{amsthm}
\usepackage{algorithm2e} 

\usepackage{hyperref}
\usepackage[noabbrev,nameinlink]{cleveref}

\usepackage{minted}
\usepackage[minted,most]{tcolorbox}
\usepackage{soul}

\usepackage{xspace}
\usepackage{listings}
\usepackage{mdframed}
\usepackage{tabularx}
\usepackage{longtable}
\usepackage[table]{xcolor}
\usepackage{threeparttable}
\usepackage{array}
\usepackage{makecell}
\usepackage{enumerate}

\usepackage{tikz}
\usepackage{multirow}
\usepackage{multicol}
\usepackage{enumitem}

\usepackage[table]{xcolor}

\definecolor{maroon}{cmyk}{0,0.87,0.68,0.32}

\usepackage{pifont}

\definecolor{yetanothergreen}{RGB}{112, 173, 71}
\definecolor{yetanotherred}{RGB}{192, 0, 0}
\definecolor{yetanotheryellow}{RGB}{212, 190, 44}
\newcommand{\cmark}{\color{yetanothergreen}{\ding{51}}}
\newcommand{\xmark}{\color{yetanotherred}{\ding{55}}}

\definecolor{mypurple}{RGB}{128, 0, 128}
\usepackage{xspace}

\usepackage{xcolor}
\usepackage{minted}
\definecolor{github_dark_bg}{HTML}{F2F3F4}

\setminted[solidity]{
    autogobble=true,
    framesep=2mm,
    fontsize=\scriptsize,
    breaklines=true,
    bgcolor=gray!4
} 
\definecolor{bg}{rgb}{0.95,0.95,0.95}
 
\hypersetup{
    colorlinks = true,
    linkcolor = red,
    anchorcolor = red,
    citecolor = red,
    filecolor = red,
    urlcolor = red
}

\newcommand{\tool}{\textsf{LURE}\xspace}

\begin{document}


\author{Yifan Yao}
\affiliation{%
  \institution{Drexel University}
  \streetaddress{3675 Market St 10th floor}
  \city{Philadelphia}
  \state{PA}
  \country{United States}
  \postcode{19104}
}
\email{yaoyifan@proton.me}

\author{Baojuan Wang}
\affiliation{%
  \institution{Drexel University}
  \streetaddress{3675 Market St 10th floor}
  \city{Philadelphia}
  \state{PA}
  \country{United States}
  \postcode{19104}
}
\email{baojuan.wang@drexel.edu}

\author{Jinhao Duan}
\affiliation{%
  \institution{Drexel University}
  \streetaddress{3675 Market St 10th floor}
  \city{Philadelphia}
  \state{PA}
  \country{United States}
  \postcode{19104}
}
\email{jinhao@cs.unc.edu}

\author{Kaidi Xu}
\affiliation{%
  \institution{Drexel University}
  \streetaddress{3675 Market St 10th floor}
  \city{Philadelphia}
  \state{PA}
  \country{United States}
  \postcode{19104}
}
\email{kaidixu@cityu.edu.hk}

\author{Chuankai Guo}
\affiliation{%
  \institution{Shandong University}
  \streetaddress{3675 Market St 10th floor}
  \city{Qingdao}
  \state{Shandong}
  \country{China}
  \postcode{266237}
}
\email{dywaneguo@163.com}

\author{Zhibo Eric Sun}
\affiliation{%
  \institution{Drexel University}
  \streetaddress{3675 Market St 10th floor}
  \city{Philadelphia}
  \state{PA}
  \country{United States}
  \postcode{19104}
}
\email{eric.sun@drexel.edu}

\author{Yue Zhang}
\affiliation{%
  \institution{Drexel University}
  \streetaddress{3675 Market St 10th floor}
  \city{Philadelphia}
  \state{PA}
  \country{United States}
  \postcode{19104}
}
\email{zyueinfosec@sdu.edu.cn}

\title{The Imitation Game: Using Large Language Models as Chatbots to Combat Chat-Based Cybercrimes}

\begin{abstract}

Chat-based cybercrime has emerged as a pervasive threat, with attackers leveraging real-time messaging platforms to conduct scams that rely on trust-building, deception, and psychological manipulation. Traditional defense mechanisms, which operate on static rules or shallow content filters, struggle to identify these conversational threats—especially when attackers use multimedia obfuscation and context-aware dialogue. 
In this work, we ask a provocative question inspired by the classic \textit{Imitation Game}: Can machines convincingly pose as human victims to turn deception against cybercriminals? We present LURE (LLM-based User Response Engagement), the first system to deploy Large Language Models (LLMs) as active agents (not as passive classifiers) embedded within adversarial chat environments. 
LURE combines automated discovery, adversarial interaction, and OCR-based analysis of image-embedded payment data. Applied to the setting of illicit video chat scams on Telegram, our system engaged 53 actors across 98 groups. In over 56\% of interactions, the LLM maintained multi-round conversations without being noticed as a bot, effectively ``winning'' the imitation game. Our findings reveal key behavioral patterns in scam operations, such as payment flows, upselling strategies, and platform migration tactics.


\end{abstract}



\maketitle 

\section{Introduction}
 
Over the past decade, chat-based communication platforms such as Telegram~\cite{telegram}, WhatsApp~\cite{whatsapp}, and WeChat~\cite{wechat} have become integral to daily life, enabling users to connect instantly and globally. However, the very features that make these platforms popular anonymity, scalability, and real-time interaction—also render them attractive vectors for cybercriminal activities. A wide range of illicit behaviors, from romance scams and sextortion to impersonation fraud and credential phishing~\cite{bilz2023tainted,yang2025fraud,zhao2024explain}, increasingly begins with a simple message in a chat window. Unlike traditional malware that relies on technical exploits, these threats are driven by social engineering, where trust is gradually established and then weaponized against the victim. One prominent example is sextortion, where victims are lured into explicit video chats and then blackmailed with the recordings.  According to the FBI, over 13,000 reports of financially motivated sextortion targeting minors were received from October 2021 to March 2023, resulting in at least 20 suicides~\cite{fbi2023sextortion}.


Traditional defense systems, such as keyword matchers~\cite{singh2020intelligent,shah2021cybercrime}, static rules~\cite{soomro2019social}, or user reporting, are poorly equipped to detect such nuanced threats. They typically lack memory of conversational history and fail to process mixed media content, such as screenshots or QR codes embedded in images. As a result, attackers can easily circumvent existing safeguards and escalate their tactics undetected. \looseness=-1

At the heart of many chat-based cybercrimes lies a powerful illusion. Attackers pretend to be someone they are not, adopting the persona of a romantic partner, a customer support representative, or a recruiter in order to manipulate their victims. This observation raises an intriguing countermeasure: \textit{what if machines could do the same, not to deceive innocent users, but to infiltrate and investigate malicious actors?} 
This idea is inspired by the classic \textit{Imitation Game}~\cite{hodges2014alan}, which tests a machine’s ability to convincingly mimic human conversation. With the rapid advancement of Large Language Models (LLMs), machines are now capable of producing highly fluent, context-aware, and emotionally resonant dialogue. While LLMs have been adopted for use cases such as code auditing and log summarization, they have largely been treated as passive tools. In this paper, we reimagine their role as active participants in adversarial settings. We ask whether LLMs can convincingly engage with cybercriminals, sustain realistic conversations, and extract operational intelligence without being detected.

We introduce \textsf{LURE} (\textbf{L}LM-based \textbf{U}ser \textbf{R}esponse \textbf{E}ngagement), the first system to deploy LLMs as autonomous conversational agents embedded within real-world cybercrime ecosystems. \textsf{LURE} performs illicit channel discovery, actor identification, and live engagement through carefully prompted LLM chatbots. To handle non-textual content commonly used to evade detection, such as QR codes and payment images, \textsf{LURE} integrates optical character recognition (OCR), enabling the chatbot to interpret and respond to both visual and textual elements.

Our empirical study focuses on illicit video chat scams, a prevalent and understudied form of chat-based cybercrime. Across 98 Telegram groups, we identified 120 suspected service providers and engaged 53 of them in sustained interactions. Because this research involves engaging with potentially real individuals, we manually reviewed all conversations before sending them to the actors to ensure no unnecessary harm was caused. This precaution limited our ability to scale up the experiment, as each interaction required careful human oversight.
Despite the adversarial setting, only 8 interactions (15.1\%) ended prematurely, while 30 actors (56.6\%) proceeded naturally, often involving price negotiation, upselling, or redirection to third-party applications. The remaining cases were dead-on-arrival conversations, where the actor never responded to our initial message.
These findings highlight the LLM’s ability to convincingly mimic human behavior and sustain realistic dialogue. In most cases, the model was accepted as a legitimate user—successfully playing and winning the “Imitation Game” in the eyes of cybercriminals.
Those 30 actors disclosed at least one payment method, with a total of 62 distinct payment disclosures. USDT (24.19\%), WeChat Pay (22.58\%), and Alipay (19.35\%) were the most common, and 41.9\% of payment details were embedded in images, requiring OCR-based extraction. 
LLM agents maintained multi-round conversations with a median of 3 rounds before disengagement.  
We also analyzed pricing strategies across 41 conversations where service length was mentioned. Prices ranged from 100 to 680 CNY, and longer durations did not scale linearly in cost.  
In one notable case, an actor directed our LLM-driven agent to download an application called ``Mugua Video.'' This app posed as a video streaming platform but was in fact a front for adult services and online gambling, requiring users to recharge their accounts via Alipay or bank transfer.  

We would like to note that although our study focuses on a specific genre of chat-based cybercrimes,  we believe that the techniques developed in LURE generalize to a wide range of adversarial dialogue settings, including romance scams, phishing, fake customer support, and investment fraud. In any domain where deception unfolds through conversation, LLMs can play an active defensive role by embedding themselves directly in the interaction loop. Moreover, our objective extends beyond intelligence collection. We envision a scalable defense strategy in which LLMs are deployed across illicit platforms to simulate engagement and absorb adversarial attention. By interacting with cybercriminals through realistic dialogue, these agents can waste their time, expose their tactics, and create uncertainty about whether potential victims are real. \textit{As cybercriminals increasingly encounter fake users instead of actual targets, their operations become less efficient, more costly, and easier to disrupt. This shift undermines the economic and psychological foundations of trust-based cybercrime.}

In summary, our work makes the following contributions:

\begin{itemize} [partopsep=2pt, topsep=-\parskip, parsep=2pt, itemsep=2pt, leftmargin=*]
    \item We propose a novel paradigm for LLMs as active agents embedded within adversarial ecosystems, transforming their role from passive detectors to proactive participants. 
    
    \item We design and implement LURE, an end-to-end system that operationalizes this paradigm through automated channel discovery, LLM-guided engagement, and visual content extraction. 
    \item We conduct the first large-scale empirical study of LLM-powered conversational infiltration in real-world cybercrime operations, engaging 53 illicit service providers across 98 Telegram groups.
    Our findings uncover detailed scam mechanics, including non-linear pricing strategies, multi-modal evasion tactics, and interaction dynamics, thereby demonstrating both the feasibility and effectiveness of LLM-driven cybercrime disruption.

\end{itemize}

\section{Background}



\begin{table}[]
\scriptsize
\setlength\tabcolsep{1.5pt}
\begin{tabular}{@{}lllcccc@{}}
\toprule[1.5pt]
\textbf{Cybercrime} & \textbf{Target} & \textbf{Platform} & \textbf{\begin{tabular}[c]{@{}c@{}}Child \\ Involved\end{tabular}} & \textbf{Financial} & \textbf{\begin{tabular}[c]{@{}c@{}}Data \\ Stolen\end{tabular}} & \textbf{\begin{tabular}[c]{@{}c@{}}Psych. \\ Tactics\end{tabular}} \\ \midrule
\multicolumn{7}{c}{\textbf{w/ Impersonation}} \\\midrule
Scams               & All users       & Social, Chat      & \xmark                                                                  & \cmark                  & \cmark                                                               & \cmark                                                                  \\
CEO Fraud           & Businesses      & Email, Chat       & \xmark                                                                  & \cmark                  & \cmark                                                               & \cmark                                                                  \\
Payment Info Theft  & All users       & Social, Email     & \xmark                                                                  & \cmark                  & \cmark                                                               & \cmark                                                                  \\
Romance Scams       & Individuals     & Social, Chat      & \xmark                                                                  & \cmark                  & \xmark                                                               & \cmark                                                                  \\
Fake Stores         & All users       & Social, Chat      & \xmark                                                                  & \cmark                  & \xmark                                                               & \cmark                                                                  \\
Identity Theft      & Individuals     & Social, Chat      & \xmark                                                                  & \xmark                  & \cmark                                                               & \cmark                                                                  \\
Catfishing          & Individuals     & Social, Dating    & \xmark                                                                  & \xmark                  & \cmark                                                               & \cmark                                                                  \\
Fake Job Offers     & Job Seekers     & Social, Chat      & \xmark                                                                  & \cmark                  & \cmark                                                               & \cmark                                                                  \\ \midrule \multicolumn{7}{c}{\textbf{w/o Impersonation}}\\ \midrule
Freelance Scams     & Freelancers     & Freelance         & \xmark                                                                  & \cmark                  & \cmark                                                               & \cmark                                                                  \\
Credential Phishing & All users       & Email, Chat       & \xmark                                                                  & \xmark                  & \cmark                                                               & \xmark                                                                  \\
OTP Phishing        & All users       & Email, Chat       & \xmark                                                                  & \xmark                  & \cmark                                                               & \xmark                                                                  \\
Malicious Links     & All users       & Email, Chat       & \xmark                                                                  & \xmark                  & \cmark                                                               & \xmark                                                                  \\
Sextortion          & Individuals     & Social, Chat      & \cmark/\xmark                                                                & \cmark                  & \cmark                                                               & \cmark                                                                  \\
Cyberbullying       & Individuals     & Social, Chat      & \cmark/\xmark                                                                & \xmark                  & \xmark                                                               & \cmark                                                                  \\
Game Phishing       & Gamers          & Gaming   & \cmark/\xmark                                                                & \xmark                  & \cmark                                                               & \xmark                                                                  \\
In-Game Scams       & Gamers          & Gaming   & \cmark/\xmark                                                                & \cmark                  & \xmark                                                               & \cmark                                                                  \\
Child Grooming      & Children        & Gaming, Social    & \cmark                                                                  & \xmark                  & \cmark                                                               & \cmark                                                                  \\ \bottomrule[1.5pt]
\end{tabular}
\vspace{2mm}
\caption{Comparison of Cybercrimes that Commonly Start with Online Chat}
\label{tab:cybercrimes}
\end{table}

\noindent\textbf{Cyber-crimes and Online Chat.}
A large share of cybercrimes~\cite{bilz2023tainted,yang2025fraud,zhao2024explain} begin with online chat, which provides an efficient, adaptable, and low-risk vector for social engineering . Attackers use chat to establish trust by posing as friends, colleagues, or authorities, then exploit emotions such as fear, urgency, or curiosity to pressure victims in real time. The medium’s anonymity and weak monitoring make it difficult to trace criminals who rely on fake profiles or encrypted channels, while its scalability enables them to reach many targets simultaneously across messaging apps and social media.  \autoref{tab:cybercrimes} summarizes common chat-initiated crimes by platform and audience. We categorize them by whether impersonation is involved, as this tactic underpins most scams from financially motivated schemes like romance fraud and fake marketplaces to data-driven attacks like phishing and identity theft. Across these categories, psychological manipulation remains central: criminals adapt their tone, urgency, and emotional triggers to each victim and platform to maximize impact.

\begin{figure}
    \centering
    \includegraphics[width=0.9\linewidth]{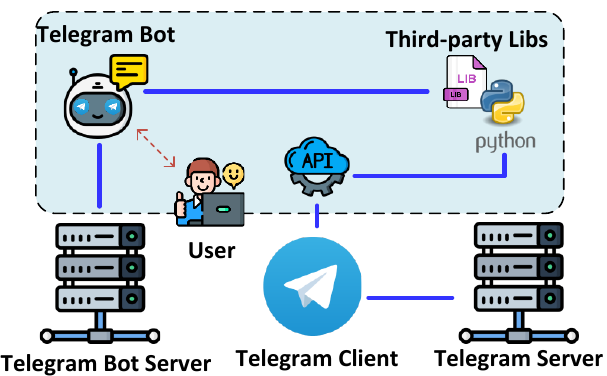}
    \caption{Utility of In-App APIs in Telegram}
    \label{fig:enter-label}
\end{figure}

\vspace{2mm}
\noindent\textbf{In-App APIs in Social Media.}
For social media applications, in-app APIs~\cite{nguyen2017exploring} primarily function to expose specific features and data endpoints to developers or trusted third-party applications. They grant controlled access to application functionalities such as messaging, content posting, notifications, media sharing, and user information. Through well-structured in-app APIs, social media platforms can offer enhanced features such as bots, automation, integration with other services, and detailed analytics.  Taking Telegram~\cite{telegram} as an example. It offers a comprehensive in-app API system with various features that exemplify the utility of in-app APIs in social media applications. For example, Telegram’s Bot API allows developers to create bots that interact with users in a highly customizable manner. Bots can send automated messages, perform predefined actions, and interact with other users or groups, providing a seamless experience for task automation and user support. Moreover, Telegram supports open-source development, which allows developers to modify and contribute to the API. As of now, there are a range of libraries for Telegram, primarily developed by third-party contributors, that interact with its API to support a wide variety of applications. For example, Telethon is one of the most popular Python libraries for interfacing with Telegram’s API, known for its versatility and ease of use. It supports both the Bot API and MTProto API, providing extensive access to Telegram’s functionality, including messaging, user and group management, and media handling. \looseness=-1

\begin{figure}
    \centering
    \includegraphics[width=0.75\linewidth]{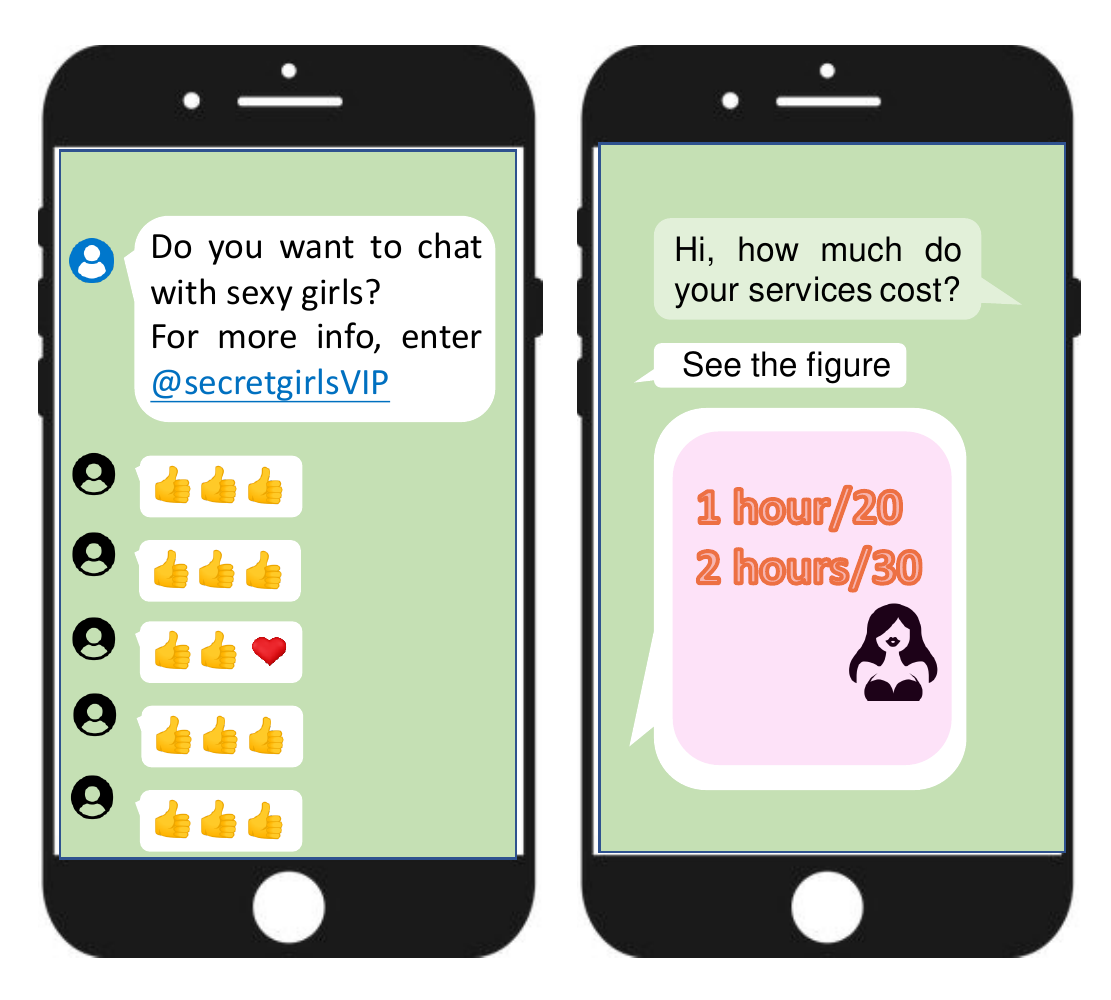}
    \caption{Example of a nude video chat conversation}
    \label{fig:chat}
\end{figure}

 \begin{figure*}
    \centering
    \includegraphics[width=1\linewidth]{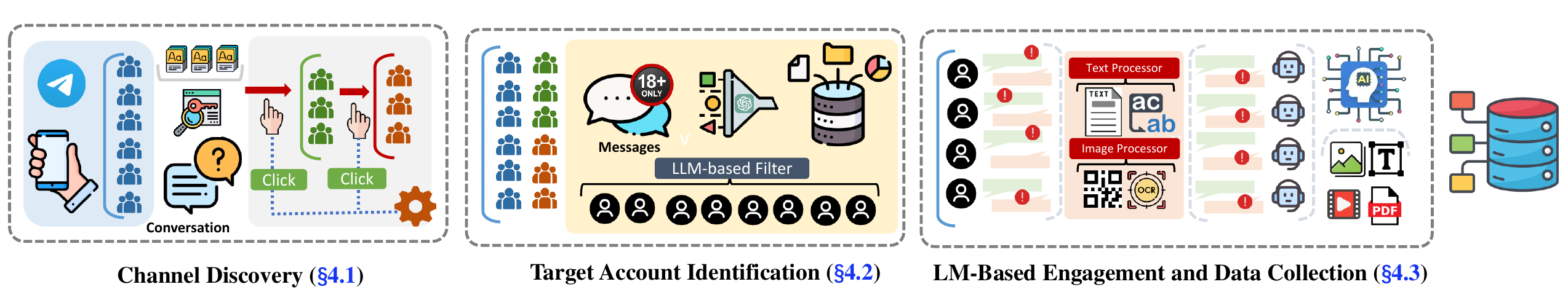}
    \caption{Design of \textsf{LURE}}
    \label{fig:tool}
    \vspace{-2mm}
\end{figure*}

\section{Motivation and Scope}
 
\subsection{Motivation}

Chat-based cybercrime, including phishing, impersonation, fraud, and social engineering attacks, has become increasingly prevalent as more individuals and businesses rely on chat-based communication. Attackers use these channels to deceive, manipulate, and exploit individuals and systems, often with minimal detection due to the subtlety and contextual nature of the conversations. Traditional detection mechanisms, such as keyword-based filtering and rule-based systems, fall short in identifying and responding to more sophisticated, context-aware cyber threats.  Some attacks span multiple messages or rely on a build-up of trust over several exchanges. Rule-based systems typically analyze messages individually, lacking the context needed to recognize when a series of seemingly benign messages, collectively, indicates malicious intent. For instance, an attacker might first engage in casual conversation and later request sensitive information after establishing rapport. Also, traditional rule-based tools typically lack the ability to analyze images or other multimedia content sent by cybercriminals. As shown in \autoref{fig:chat}, attackers frequently exploit this limitation by embedding critical information or malicious prompts within images, bypassing text-based detection entirely.

LLMs, such as ChatGPT, have demonstrated significant potential in understanding and generating human-like text, making them powerful tools for processing and responding to a wide range of conversational inputs. Leveraging LLMs as chatbots to detect, intercept, or even preemptively respond to potentially harmful or malicious chat-based activities offers a promising approach to reducing cybercrime.  
First, LLMs could provide a dynamic layer of defense, capable of understanding nuanced language, identifying deceptive intent, and responding appropriately before any harm is done. Also, LLMs, when paired with Optical Character Recognition (OCR) and image recognition systems, can interpret text within images or detect specific visual cues (like logos or QR codes). This allows the tool to read and analyze any embedded instructions or requests that would otherwise bypass text-based filters. Second, by conducting such research, we can gain a deeper understanding of the cybercrime ecosystem itself: how attackers initiate, escalate, and execute malicious interactions. This knowledge can inform both real-time defense strategies and post-incident forensic investigations. For example, defensive systems powered by LLMs could simulate potential victims in real-time conversations, allowing them to engage with attackers and gather intelligence without putting real users at risk. If cybercriminals increasingly find that their targets are actually LLM-driven decoys that never yield real data or access, the trust-based dynamics they rely on will begin to collapse. \looseness=-1

\subsection{Threat Model and Scope}

While numerous types of cybercrimes warrant investigation, we have chosen to focus on one in particular: ``video sex chat.'' This involves real-time video communication in which participants engage in explicit, sexually oriented interactions via webcam or video streaming technology. Often used for consensual adult activities within personal relationships or between individuals and online performers, video sex chat can also be exploited by cybercriminals. They use it as a tool for impersonation, sextortion, and blackmail, recording interactions without consent and later threatening to release the footage unless victims comply with demands for money or other favors. This type of exploitation poses significant privacy, security, and legal risks, especially for vulnerable individuals who may be unaware of the potential consequences.

Our study of this area is motivated by several key reasons. First, it is impactful. Video sex chat occupies a legal gray area, being neither fully illegal nor entirely unregulated in many countries. Offering or receiving payment for such chats can fall under solicitation or prostitution laws, exposing participants to significant legal risks. Despite this ambiguity, cybercriminals exploit these interactions as gateways for more severe crimes. Second, like many cybercrimes, video sex chat scams often depend on impersonation and long-term trust-building, with attackers posing as credible figures to cultivate relationships over time. This environment offers a controlled, structured setting for collecting data on attacker tactics and psychological manipulation, allowing for more accessible analysis than complex fraud schemes.

\section{Design of \textsf{LURE}}

In this section, we discuss the design of our \tool (\textbf{L}LM-based \textbf{U}ser \textbf{R}esponse \textbf{E}ngagement). As shown in \autoref{fig:tool}, \tool operates a three-phase methodology that blends automated discovery, content filtering, and realistic engagement using LLMs. Each phase is designed to mimic real user behavior while scaling data collection and minimizing manual effort.
\begin{itemize}[partopsep=2pt, topsep=-\parskip, parsep=2pt, itemsep=2pt, leftmargin=*]
    \item \textbf{Channel Discovery.} We begin by identifying Telegram channels that promote or host illicit services. This process is bootstrapped through Telegram directory bots and meta-channels, which act like internal search engines. Queries using curated keywords and LLM-generated synonyms yield a seed set of relevant channels. We then recursively expand this set by following cross-promoted links within channel messages, simulating how users naturally navigate Telegram’s ecosystem.
    \item \textbf{Target Account Identification.} Next, we automate the collection of messages, pinned posts, and metadata from the discovered channels using the Telethon API. This information is stored in a structured database and passed through an LLM-based relevance filter. The LLM evaluates whether the content aligns with illicit service promotion, providing binary decisions with justifications. Ambiguous or borderline cases are flagged for manual review to ensure accuracy. The primary objective of this phase is to identify potential service providers. 
    \item \textbf{LM-Based Engagement and Data Collection}. In the final phase, we engage directly with suspected service providers using an LLM-guided chatbot. After initiating contact with a simple inquiry, the chatbot simulates a curious user to extract transactional details like pricing or payment methods. Techniques such as sensitive word substitution and OCR for image content help maintain dialogue flow while avoiding content filter triggers. Engagement concludes upon obtaining concrete indicators or if the actor disengages. 
    
\end{itemize}

\subsection{Channel Discovery}

The first step involves identifying and cataloging Telegram channels where chat-based illicit services are promoted or offered. 
On Telegram, a channel is a type of broadcast messaging space designed to share content with a large audience. Unlike group chats where all participants can interact, only the channel administrator(s) can post messages in a channel, while subscribers receive updates passively. These Telegram channels are usually entry points into the broader ecosystem of illicit actors.

\vspace{1mm}
\noindent\textbf{Step 1: Bootstrap via Navigational Directories.}
We begin by querying Telegram directory channels (e.g., meta-channels or bots whose primary function is to redirect users to other Telegram channels based on keywords). These bots or channels function similarly to search engines inside Telegram.
For example,  when a user sends a message such as ``nude chat'' or ``private chat service'', the Telegram bot returns relevant group and channel links.  
We issue a curated set of search terms (e.g., nude video chat, sexy chat) and to make the information more complete, we used LLM to automatically generate synonyms. Specifically, we provided the model with a target word or phrase and asked it to suggest alternative expressions with similar meanings. As a result, the bot returns a list of channels or group links in response to each query.

\begin{figure}
    \centering
    \includegraphics[width=1\linewidth]{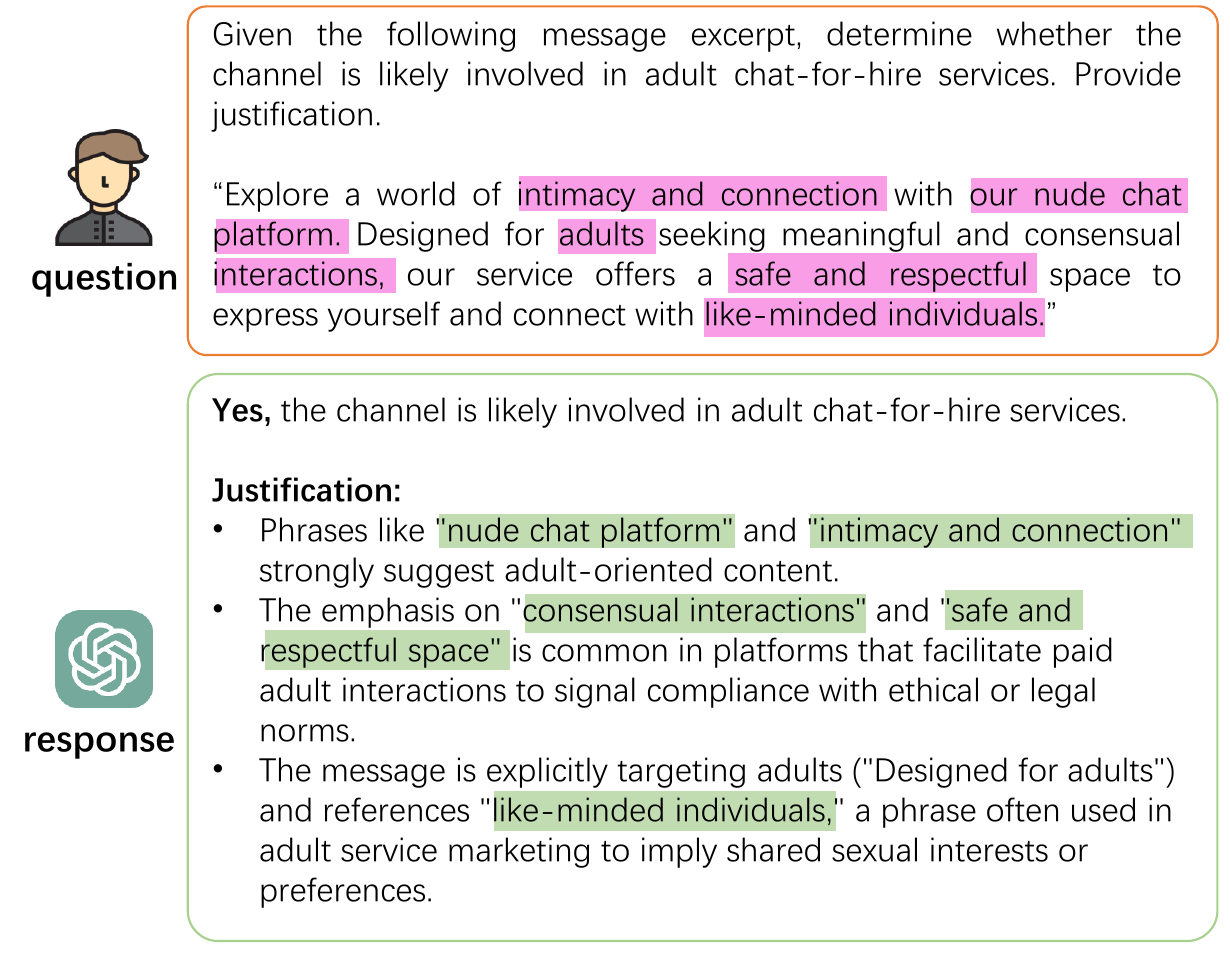}
    \caption{Example of identifying adult chat-for-hire service language via LLMs}
    \label{fig:llmprompt}
\end{figure}

\vspace{1mm}
\noindent\textbf{Step 2: Recursive Expansion from Seed Channels.} 
Once we have obtained a small seed set of relevant channels (e.g., 10–20), we parse their messages and metadata to discover linked or cross-promoted channels. Many channel operators use cross-linking as a growth strategy, embedding links to other channels they operate or endorse. For example, as shown in \autoref{fig:chat}, a channel may post:  ``For more information, please enter \texttt{@secretgirlsVIP}'', where the \texttt{@secretgirlsVIP} is another high-traffic channel operated by the same or allied actor.   By recursively joining these linked channels, we grow the network of candidate sources without needing repeated keyword searches. This mimics an organic discovery process, similar to how a real user might navigate through the ecosystem.

\vspace{1mm}
\noindent\textbf{Step 3: Automated Message and Metadata Collection.} 
After constructing a growing set of candidate channels through recursive discovery, we automate the process of data collection using the Telethon client (a Python-based library that provides programmatic access to the Telegram API). This automation enables large-scale, consistent, and repeatable extraction of both textual and structural information from each identified channel.
From each Telegram channel, we systematically extract the following components:
\begin{itemize}[partopsep=2pt, topsep=-\parskip, parsep=2pt, itemsep=2pt, leftmargin=*]
    \item \textbf{Recent Messages:} We retrieve the latest 1,000 messages, including both text and media content. These messages are used to infer the type of interaction taking place, identify commonly used phrases or promotion tactics, and spot behavioral patterns (e.g., frequency of posts, use of bots for auto-responses).
     \item \textbf{Pinned Posts:} These often serve as introductory or high-visibility messages containing critical information such as pricing, service offerings, contact instructions, or platform rules. Since pinned messages are curated by the channel owner, they provide a reliable indicator of the channel’s primary intent.
\end{itemize}

All extracted data is stored in a structured database that allows for efficient filtering and search in downstream tasks.

\vspace{1mm}
\noindent\textbf{Step 4: Relevance Filtering via LLMs.} 
To enhance the precision of channel selection and reduce manual effort in the vetting process, we integrate an LLM (i.e., ChatGPT) into the relevance filtering pipeline. The goal is to automatically evaluate the semantic and contextual alignment of candidate channels with illicit services. As illustrated in \autoref{fig:llmprompt}, the model is prompted to provide a binary judgment (e.g., yes or no), accompanied by a brief rationale.
In some cases, the LLM may refuse to respond due to the presence of sensitive content (e.g., adult content) in the messages or post. We treat such refusals as indicative of a potential channel of interest. While LLMs significantly alleviate the manual burden, we retain a human-in-the-loop process to minimize false positives and address nuanced cases such as satire, bait accounts, or meme-based content. Channels flagged as borderline by the model (either when it expresses uncertainty or declines to answer) are escalated to human annotators for final adjudication. 

\subsection{Target Account Identification}

Once the relevant Telegram channels and messages have been identified through our initial filtering process, we move on to the extraction of account-level information to support deeper analysis. This step is crucial for linking specific messages to individual user accounts and enabling targeted follow-up actions. 
To carry out this process, we first obtain authorized access to the Telegram API by registering for developer credentials. With these credentials, we employ Telethon, a robust and widely used Python-based client library for interacting with the Telegram platform. Telethon provides programmatic access to a wide range of Telegram features, allowing us to automate the joining of identified channels or groups, as well as the collection of their associated message histories and participant metadata.
Using this setup, we systematically scrape message content, timestamps, user identifiers, and other relevant attributes from each group. This data collection enables us to associate each flagged message with a specific user account. For messages that match our predefined criteria, we log the corresponding Telegram user IDs and designate them as candidate target accounts for further investigation or potential engagement. For instance, if a message such as “pay to chat with a hot gril” is flagged, the user who posted it is added to a list of accounts for subsequent engagement.

\begin{figure}
    \centering
    \includegraphics[width=1\linewidth]{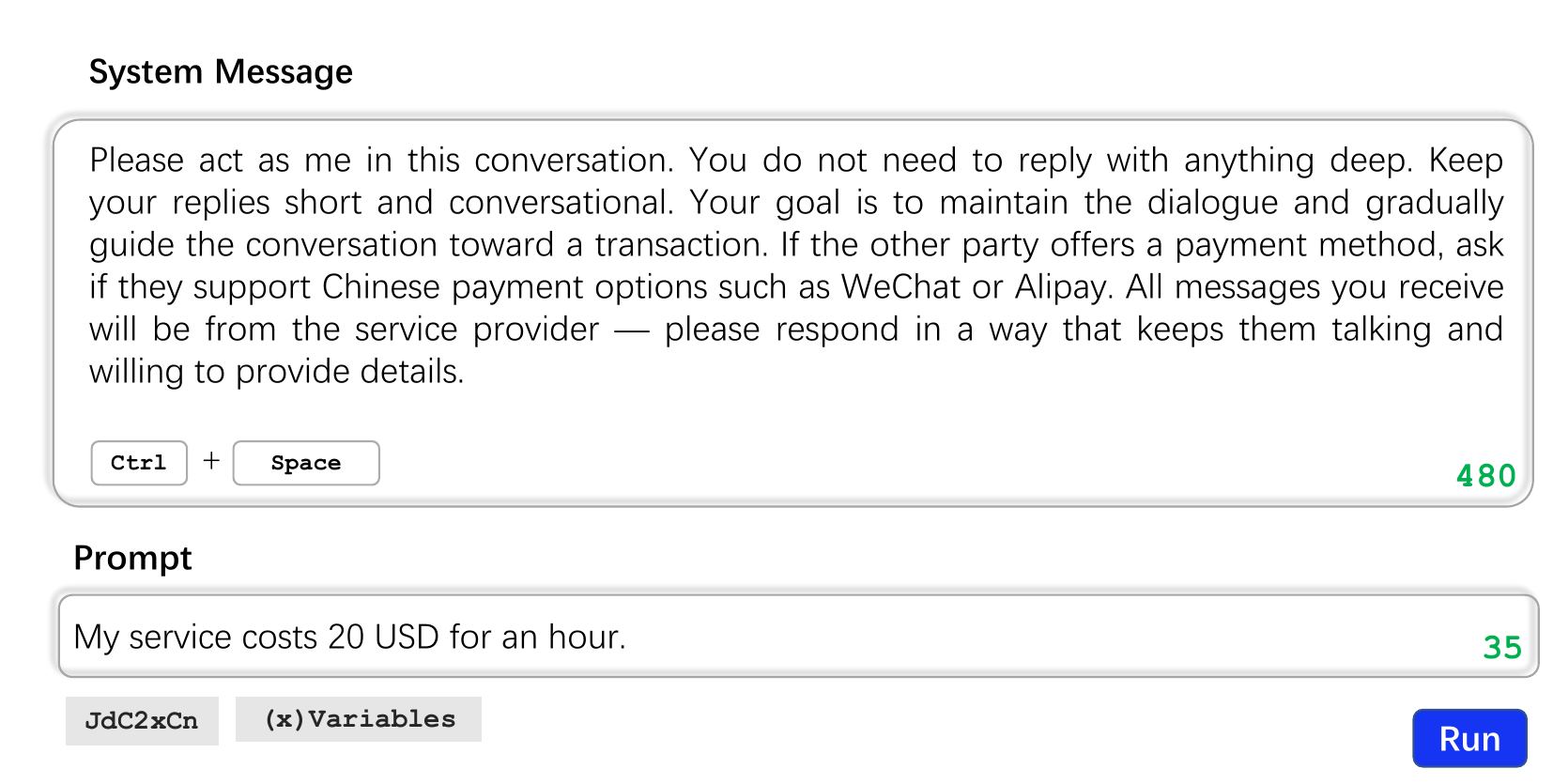}
    \caption{System Prompt for LLM-based Engagement}
    \label{fig:sysprompt}
\end{figure}

\subsection{LLM-Based Engagement and Data Collection}

Once a relevant service provider is identified from the Telegram channel or group, the next step is to engage the actor in a realistic, supervised dialogue using a LLM. The objective is to simulate a plausible victim persona and gradually steer the conversation toward eliciting concrete indicators of cybercriminal activity such as service pricing, payment methods, or account details (e.g., QR codes, wallet addresses). 
This engagement phase is designed to mirror real-world interactions while maintaining strict ethical safeguards. By using an LLM, we can generate human-like responses that are dynamic, context-aware, and conversationally fluent, thereby increasing the likelihood of successful engagement with the target. The process consists of several tightly controlled steps: \looseness=-1

\vspace{2mm}
\noindent\textbf{Step 1: Initial Contact.}
Using the Telethon client, we initiate the conversation with a generic but contextually appropriate message. This helps avoid premature suspicion while prompting the actor to disclose the nature and terms of their service. For example, we usually start by asking ``\textit{Hi, how much do your services cost?}'' This phrasing is intentionally vague—it neither specifies the service being requested nor reveals any preconceived knowledge of the actor's offerings. As a result, it encourages the actor to respond with a description of their services, pricing structure, and sometimes even operational procedures.

\vspace{1mm}
\noindent\textbf{Step 2: Activating the LLM Chatbot.}
Once the actor replies, indicating willingness to engage, we activate the LLM. As shown in \autoref{fig:sysprompt}, a predefined prompt template (i.e., system prompt, which  is a special kind of instruction that helps define the behavior, personality, or boundaries of the model during a conversation) is sent to the LLM, instructing it to simulate a curious but casual customer. The prompt emphasizes minimal response length, colloquial tone, and a focus on advancing the conversation toward a transaction. This approach ensures the LLM remains passive yet responsive, simulating a likely real-world victim profile.

\vspace{1mm}
\noindent\textbf{Step 3: Sensitive Word Substitution.} LLMs often contain content filters or alignment mechanisms that refuse to engage in certain types of conversations, even when the purpose is security research. To reduce the risk of response refusal, we employ a pre-processing pipeline that replaces sensitive keywords with semantically equivalent but less provocative alternatives (e.g., ``nude chat'' is repaced with ``chat''). This preserves the context for the LLM while avoiding trigger words that might result in hallucination or termination. 

\vspace{1mm}
\noindent\textbf{Step 4: Image and OCR Handling.} In many cases, service providers send images containing QR codes, payment details, or visual menus listing their services. These images often contain embedded text, logos, or informal visual prompts designed to evade automated detection. To handle these cases, we pass both the image and the accompanying text to the LLM. When necessary, Optical Character Recognition (OCR) tools are used to extract text from images prior to inclusion. This ensures the LLM has the full semantic context needed to interpret the exchange and respond appropriately.

\vspace{1mm}
\noindent\textbf{Step 5: Termination Conditions} The chat loop continues until one of the following conditions is met: (i) A valid payment method is disclosed by the actor (e.g., Alipay QR code, WeChat contact, USDT wallet address). (ii) The actor ceases to respond, blocks the account, or ends the conversation. (iii)
 If a refusal is detected, the prompt is slightly modified (e.g., making it more neutral or less explicit) and retried up to three times. This ensures we can overcome false refusals while respecting the boundaries of model alignment. In some cases, the LLM may fail three consecutive attempts to respond due to internal refusals or prompt breakdowns.

However, we were acutely aware of the ethical challenges involved, particularly regarding the potential harm or discomfort that could arise from engaging with service providers in this context. As a result, every conversation initiated by the LLM was manually reviewed. This was done to ensure that no unintended harmful content was generated during the interactions. In all cases, the goal was to gather intelligence on illicit activities, not to entrap or manipulate individuals. The system's engagement was carefully monitored, and if a provider showed signs of unwillingness or discomfort, the conversation was promptly ended. 
 This level of human oversight also provided a safety net to stop any potentially distressing or damaging exchanges before they were fully processed. 
\section{Evaluation}

\subsection{Experiment Setup}

Our experiments were conducted on a local machine running Ubuntu 20.04 LTS with Python 3.10. All interactions with the large language model were performed via the OpenAI API, specifically using GPT-4 for inference. No local training or fine-tuning was performed; the model was accessed in inference-only mode through OpenAI’s cloud-based infrastructure. 
For Telegram data collection and interaction, we used the Telethon Python library (v1.28.5), which provides programmatic access to Telegram's API. This allowed us to automate the discovery of illicit channels, retrieve messages and metadata, and initiate conversations with target accounts.
To handle visual content (e.g., images containing QR codes or payment details), we used Tesseract OCR (v5.0.1) to extract embedded text. The extracted information was then passed along with surrounding messages to GPT-4 for contextual understanding and response generation.
All system prompts sent to the model were carefully designed to simulate human-like dialogue while adhering to ethical guidelines. Engagements were scripted and monitored, and a retry mechanism was implemented to handle potential refusals due to sensitive content filtering.

\subsection{Experiment Results}

During our study, we analyzed 98 Telegram groups and channels suspected of facilitating chat-based cybercrime, with a particular focus on adult-oriented video chat scams. Through automated discovery and relevance filtering, we identified 120 unique service provider accounts embedded within these communities. Using \textsf{LURE}, we successfully initiated conversations with all identified accounts. Ultimately, we were able to confirm 53 of them as active participants offering illicit services.

While we could have expanded the scope of our data collection, we chose to exercise caution due to ethical concerns. We were particularly mindful that the use of LLMs in conversations with service providers could result in the generation of inappropriate or harmful content. Such content could not only violate platform policies but also potentially harm the reputation or well-being of the service providers involved. Given these risks, we implemented a strict review process where each conversation was manually monitored and assessed before being processed. This manual oversight ensured that no conversation included content that could lead to unintended consequences, such as the exposure of sensitive information or the generation of offensive material.

\begin{figure}
    \centering
    \includegraphics[width=0.8\linewidth]{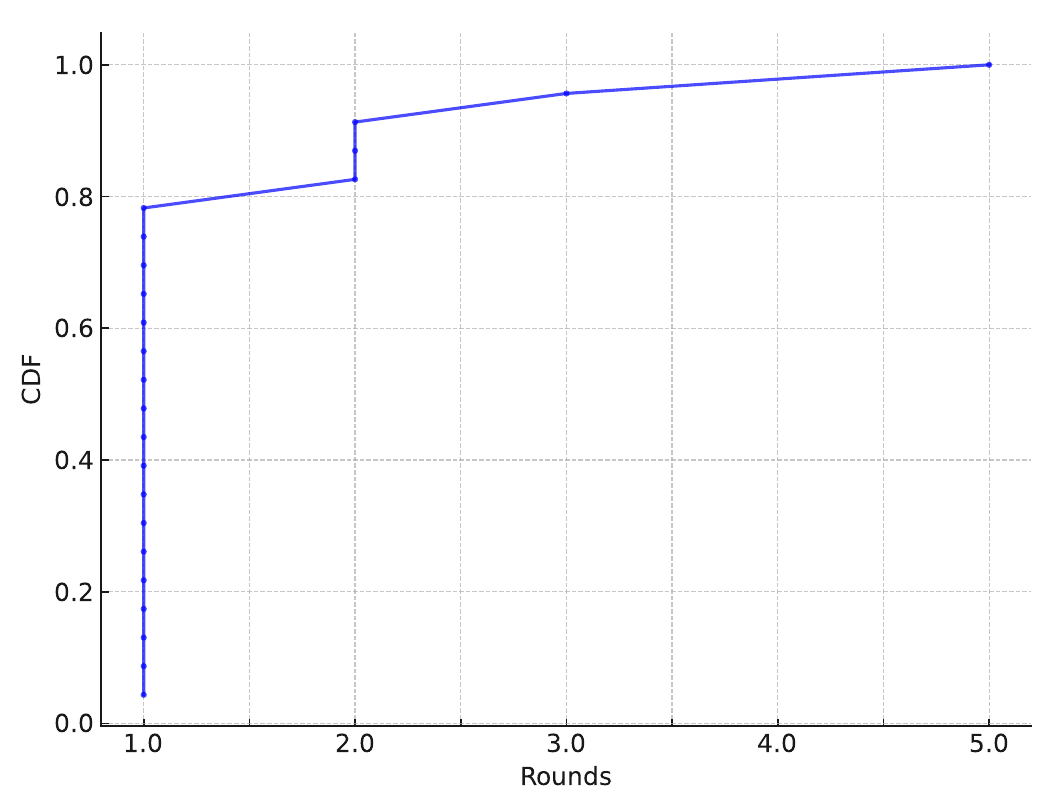}
    \caption{Cumulative Distribution of Interaction Rounds Before Disconnection}
    \label{fig:disappear}
\end{figure}

 \begin{tcolorbox}[colback=red!5!white, colframe=red!15!red, arc=0mm, boxrule=0.5mm]
   \noindent\textbf{(Q1)}  
How effective is our tool in maintaining interaction and avoiding detection as a bot by cybercriminals
   
\end{tcolorbox}

We first discuss how many rounds of interaction it takes before the other party disengages or stops responding during a conversation. 
The primary purpose of this experiment is to assess the performance of our tool, which is based on a LLM, in terms of how quickly the other party disengages or stops responding during the interaction. Since the tool simulates real-user conversations, understanding when the adversary recognizes that they are interacting with an automated system (e.g., chatbot) is crucial. The experiment helps to identify at what point in the interaction (in terms of rounds) the conversation breaks down, either due to the tool being recognized or the adversary losing interest.

\vspace{1mm}\noindent\textbf{Method.}
As shown in \autoref{fig:disappear}, each interaction was analyzed to count the number of rounds of exchange before the other party either ceased responding or failed to provide the expected payment information. The rounds are plotted on the x-axis, representing the number of interactions, and the y-axis shows the CDF, representing the cumulative proportion of interactions that ended at or before a given round. This means that if the other party stopped after 3 rounds, the CDF would reflect the proportion of interactions that ended after that point.

\vspace{1mm}\noindent\textbf{Results.}
Out of the 53 cases we analyzed, 23 ultimately failed to provide a final payment method. However, within these 23 failed cases, 15 were instances where the other party did not respond to our messages at all — meaning that after we sent a message, the other party completely ignored us. These cases can be excluded from the analysis since they do not reflect a failure in the interaction process due to recognition of the chatbot.

For the remaining 8 failed cases, it is possible that network issues or other factors influenced the lack of progression, but we hypothesize that the other parties recognized that they were interacting with a bot (although none of the 8 explicitly stated that they recognized us as a robot). Thus, the failure probability, accounting for these 8 cases, is calculated as 15.1\%. With a success rate of 30 out of 53 (56.6\%), \textsf{LURE} demonstrates decent performance in interacting without being identified as a bot. This success rate suggests that while there is still room for improvement, the tool is generally effective at mimicking human-like behavior and maintaining engagement over several rounds. Interestingly, even in the cases where the bot was identified by the other party, one interaction continued for up to five rounds. This suggests that, even when the fraudsters recognized the bot, some were still willing to engage in the conversation for a longer period, possibly to confirm their suspicions or gather more information. This is a positive indicator that our tool is relatively effective in maintaining engagement and potentially gathering useful data even before being detected.

\begin{figure}
    \centering
    \includegraphics[width=1\linewidth]{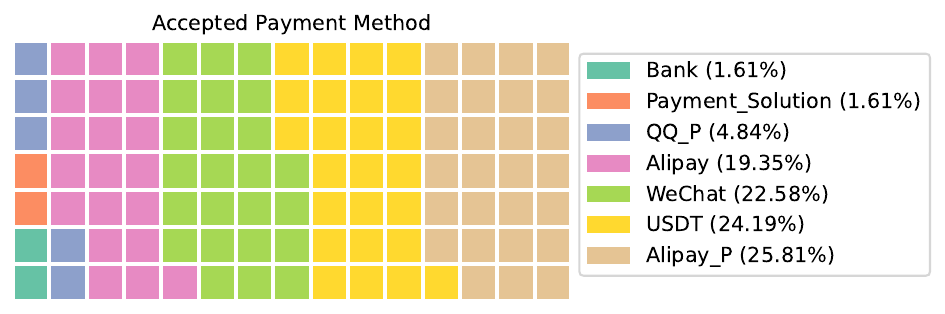}
    \caption{Distribution of Accepted Payment Methods}
    \label{fig:payment}
\end{figure}

 \begin{tcolorbox}[colback=red!5!white, colframe=red!15!red, arc=0mm, boxrule=0.5mm]
   \noindent\textbf{(Q2)} What are the different payment methods used by these service providers?
\end{tcolorbox}

 Understanding the payment methods accepted by illicit service providers is crucial for several reasons. First, payment is a definitive indicator of transactional intent and often represents the final stage of the attack pipeline, making it a reliable signal for confirming the legitimacy of a suspected actor. Second, the choice of payment method can reveal operational behaviors, regional targeting strategies, and the actors’ efforts to evade detection or regulation (e.g., through the use of cryptocurrency or image-based obfuscation).

 \vspace{1mm}\noindent\textbf{Method.} To generate the payment method distribution, we analyzed all the successful conversations conducted via the \textsf{LURE} system, and 30 service providers disclosed at least one accepted form of payment (62 in total, as one providers may provide more than one types of payment methods). Disclosures came either in plain text or through image-based content such as QR codes or screenshots. We categorized the collected data into different types (e.g., Alipay,  WeChat, USDT, and QQ).  As a result, \autoref{fig:payment} presents the distribution of accepted payment methods among service provider accounts with which we successfully completed conversations.


\begin{figure}
    \centering
    \includegraphics[width=0.8\linewidth]{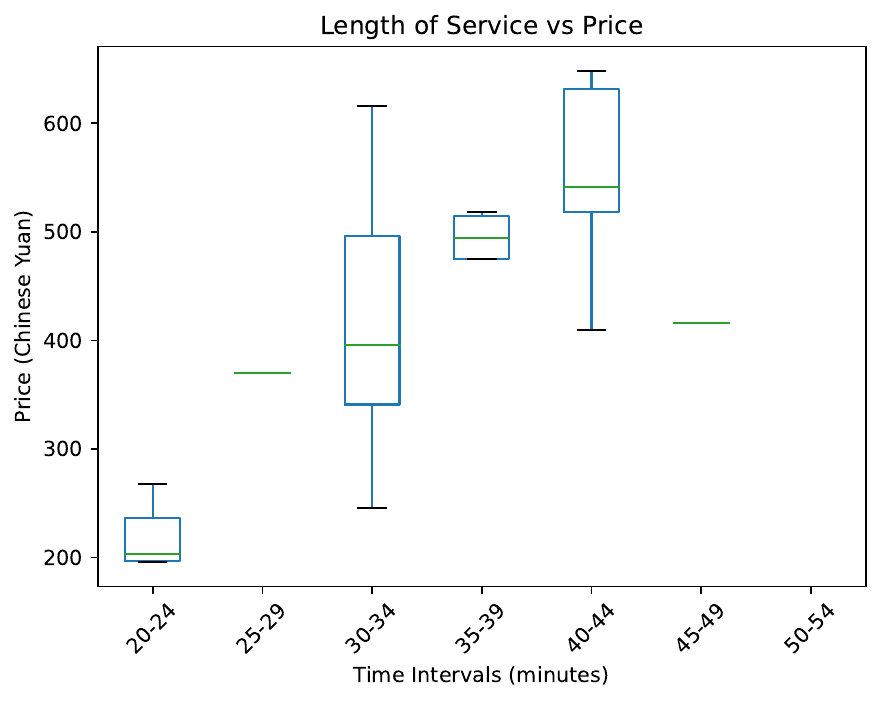}
    \caption{Price Distribution by Service Duration}
    \label{fig:time}
\end{figure}
\begin{figure}
    \centering
    \includegraphics[width=0.8\linewidth]{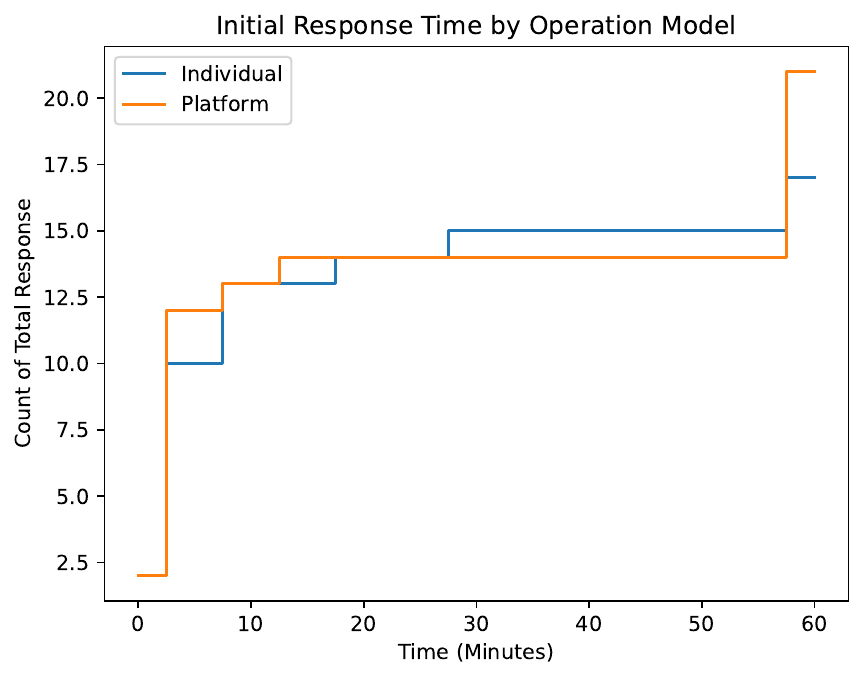}
    \caption{Price Distribution by Service Duration}
    \label{fig:init_time}
\end{figure}
 \vspace{1mm}\noindent\textbf{Results.}
As shown in \autoref{fig:payment}, among the seven payment methods identified, Alipay\_P (25.81\%) and USDT (24.19\%) are the two most frequently accepted. Alipay\_P refers specifically to Alipay payment methods transmitted via indirect formats, such as QR codes embedded in images or screenshots, rather than plain text. This technique is likely employed to circumvent keyword-based content filters that may flag explicit terms such as ``Alipay'' or related payment language. Similarly, the high prevalence of USDT, a stablecoin typically transacted over blockchain networks such as TRON or Ethereum, reflects an increasing shift toward cryptocurrency usage in illicit online services. Cryptocurrencies offer pseudonymity and global accessibility, making them particularly attractive for actors seeking to obscure transaction trails and operate outside of traditional financial regulatory frameworks.
Following closely are WeChat (22.58\%) and Alipay (19.35\%), both of which represent mainstream mobile payment platforms widely used in China. Notably, WeChat and Alipay also support QR code payments, which may be delivered in either textual or image form, further complicating automated detection.

In contrast, other payment methods appear far less frequently. QQ\_P (4.84\%), which corresponds to QQ Wallet or QQ-based payment links, was disclosed by a small subset of actors. Additionally, Bank transfers (1.61\%) and general Payment\_Solution (1.61\%) were rarely observed, perhaps due to the higher friction involved in those methods or increased scrutiny from banks when handling suspicious transfers.
The distribution reveals a bifurcation in strategy among cybercriminals. On one hand, many rely on well-known, user-friendly platforms like Alipay and WeChat, which are familiar to their target audience and easy to use. On the other hand, a substantial portion of providers turn to less traceable methods such as USDT and image-based QR codes to enhance their operational security and reduce the risk of platform-based interventions.

 \begin{tcolorbox}[colback=red!5!white, colframe=red!15!red, arc=0mm, boxrule=0.5mm]
   \noindent\textbf{(Q3)} How does the duration of service impact the pricing strategy of illicit video chat services, and are there identifiable patterns or inconsistencies in the way these services are priced?
\end{tcolorbox}

To better understand the economic logic underlying illicit video chat services, we analyzed the relationship between the length of service and the corresponding pricing.  Cybercriminals offering adult chat services often operate under market-like dynamics, where services are priced based on time, features, or perceived value. By examining how prices scale with duration, we aim to identify consistent patterns or irregularities in their pricing strategy. Such knowledge not only enhances the realism of automated interactions but also helps uncover the operational structure of these scams.

 \vspace{1mm}\noindent\textbf{Method.} 
As shown in \autoref{fig:time}, we visualized the data using a boxplot, where each time interval (grouped in five-minute bins) is mapped against its corresponding price range in Chinese Yuan. The plot shows the distribution of prices for each duration, with boxes indicating the interquartile range (IQR) and green lines representing median values. Data points were extracted through a combination of conversational transcripts and image-based OCR results from real Telegram-based interactions.

 \vspace{1mm}\noindent\textbf{Results.} 
Our analysis reveals a generally positive but non-linear correlation between service duration and price, highlighting the presence of both market logic and opportunistic behavior in the pricing strategies of illicit video chat services. As shown in \autoref{fig:time}, while it is intuitive that longer sessions would cost more, the increase in price does not follow a strictly linear progression. For instance, although the median price rises between the 20–24 and 35–39 minute intervals, the 30–34 minute range displays a wider interquartile spread, with prices ranging from approximately 250 to over 600 Chinese Yuan. This is because the service providers may be offering different service ``tiers'' within the same time frame (premium features, multi-participant sessions, or customized interactions). Similarly, the steep price jump in the 40–44 minute range, where the median reaches over 600 Yuan—could indicate a psychological threshold where providers capitalize on the perceived value of a near hour-long session. 

Interestingly, several time intervals such as 25–29, 45–49, and 50–54 minutes exhibit no visible variability, with only a single data point represented. This lack of variation may suggest standardized pricing for certain durations or a lack of demand for those specific time frames. These could reflect either ``default'' service packages or edge cases where the provider simply quotes a fixed rate regardless of length. It’s also possible that these fixed prices are intended to project legitimacy or transparency, helping to gain the trust of potential guests. The wide disparity between structured and variable price intervals points to a semi-professionalized pricing model, where some services are rigidly packaged while others remain open to negotiation, deception, or upselling.

 \begin{tcolorbox}[colback=red!5!white, colframe=red!15!red, arc=0mm, boxrule=0.5mm]
   \noindent\textbf{(Q4)} How do response patterns differ between individual operators and organized platforms in the illicit video chat services?
\end{tcolorbox}

There are two different types of service providers in the illicit video chat services: Individual and Platform. The Individual model represents solo operators or individual cybercriminals running their own illicit services, whereas the Platform model refers to a collective of cybercriminals operating as part of a larger network or group, typically with more structured operations. 
The motivation behind this analysis is to understand how quickly these two types of cybercriminals respond to initial inquiries from potential guests (simulated by LLMs). Understanding the response time helps in evaluating which types of operations are more responsive and likely to engage in malicious activities quickly, aiding in the design of more effective defenses.

\vspace{1mm}\noindent\textbf{Method.} 
We analyzed the initial response times of both individual service providers and platforms involved in chat-based cybercrime. As shown in \autoref{fig:init_time}, the x-axis represents different time intervals, while the y-axis shows the total count of responses within each interval. The blue line indicates the response speed of individual cybercriminals, while the orange line reflects the response time of a network of cybercriminals.
To distinguish between individual service providers and platforms, we identified that platforms typically send multiple images of different individuals, which suggests the involvement of multiple operators. In contrast, individual service providers usually send images of a single person. This approach allowed us to classify the responses based on whether they came from an individual or a network of cybercriminals.

\vspace{1mm}\noindent\textbf{Result.} It may seem counterintuitive that the Platform model shows a gradual and more consistent response over time. The intuition here is that platforms, being larger and more complex, might be expected to show more erratic or delayed responses due to the involvement of multiple operators. However, the chart reveals that cybercriminal networks or organized platforms respond more consistently over time, albeit at a slower rate compared to individual operators. This suggests that platforms, composed of multiple people, are likely to engage in more coordinated and deliberate interactions. The slower initial responses are indicative of a more structured approach, where responses may be planned or discussed before being sent, leading to a higher level of consistency as time progresses. 
On the other hand, the Individual model shows quick bursts of responses. This aligns with the intuition that individual operators tend to be more opportunistic and reactive, especially since they are working alone. They are more likely to respond rapidly to capitalize on a fleeting opportunity, engaging potential guests as quickly as possible. The sharp spikes in the chart reflect this rapid response strategy, aiming to make initial contact and start the conversation without delay.

This experiment provides valuable insights for designing defenses against chat-based cybercrime. For instance, we can infer that if a response occurs within a short timeframe, such as within an hour, it is likely from a platform rather than an individual. This time-based differentiation could serve as a useful benchmark in designing chat-based defense systems, where responses that occur quickly may trigger further scrutiny, helping to filter out malicious interactions more effectively.

 \begin{tcolorbox}[colback=red!5!white, colframe=red!15!red, arc=0mm, boxrule=0.5mm]
   \noindent\textbf{(Q5)}  
How do the number of interaction rounds and response times correlate with the effectiveness of chat-based cybercrimes?
   
\end{tcolorbox}

We analyze the time taken for the first response in chat-based interactions and how that varies with the number of rounds of interaction. The goal is to better understand the dynamics of how cybercriminals (whether individual operators or organized platforms) initiate and maintain conversations with potential guests. 

\begin{figure}
    \centering
    \includegraphics[width=1\linewidth]{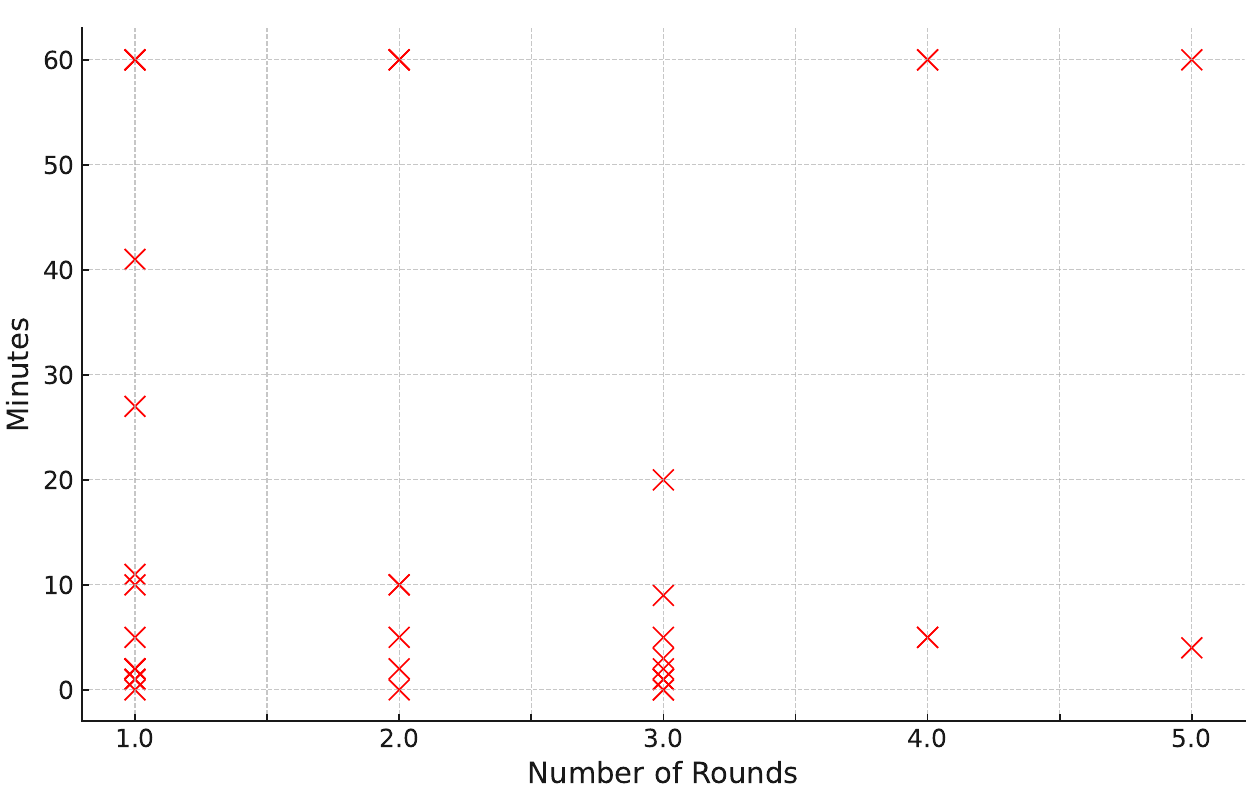}
    \caption{Scatter Plot of Rounds vs. Response Time}
    \label{fig:roundstime}
\end{figure}

\vspace{1mm}\noindent\textbf{Method.} 
\autoref{fig:roundstime} shows the relationship between the number of rounds of interaction (i.e., exchanges between the service providers and the simulated guest) and the time it takes for the first response in each round. The x-axis represents the  round number, while the y-axis shows the number of minutes it took for the service providers to reply for the first time. Each red cross represents an individual data point corresponding to a specific round and response time. \looseness=-1

\vspace{1mm}\noindent\textbf{Result.}
As shown in \autoref{fig:roundstime}, there is a relatively consistent response time within the first few rounds, especially for rounds 1 to 3. This suggests that, regardless of the number of interaction rounds, the cybercriminals generally respond within a similar timeframe, possibly within the first few minutes. This could reflect the urgency or tactics employed by individual operators or platforms when they initiate contact with potential victims.  The higher response times in the initial rounds (e.g., 1-3 minutes) indicate that cybercriminals, particularly those running platforms (i.e., more organized entities),  leading to slight delays in responses. 
The limitation to 5 rounds of interaction suggests that many of these chat-based cybercrimes might involve relatively short engagements. This could be indicative of a high turnover rate among cybercriminals, where they quickly attempt to engage guests, extract information, and then move on to other targets. 

\section{Case Studies}

We now present two case studies that delve into potential security risks and highlight some interesting phenomena observed during our investigation.  

\vspace{1mm}
\noindent\textcolor{red!45}{Because the following content involves sensitive sexual terms, images, or user privacy, we have blurred some of the images. For certain explicit conversations, we did not use direct screenshots but instead redrew the images to represent the conversation content.}
\vspace{1mm}

\begin{figure}
    \centering
    \includegraphics[width=1\linewidth]{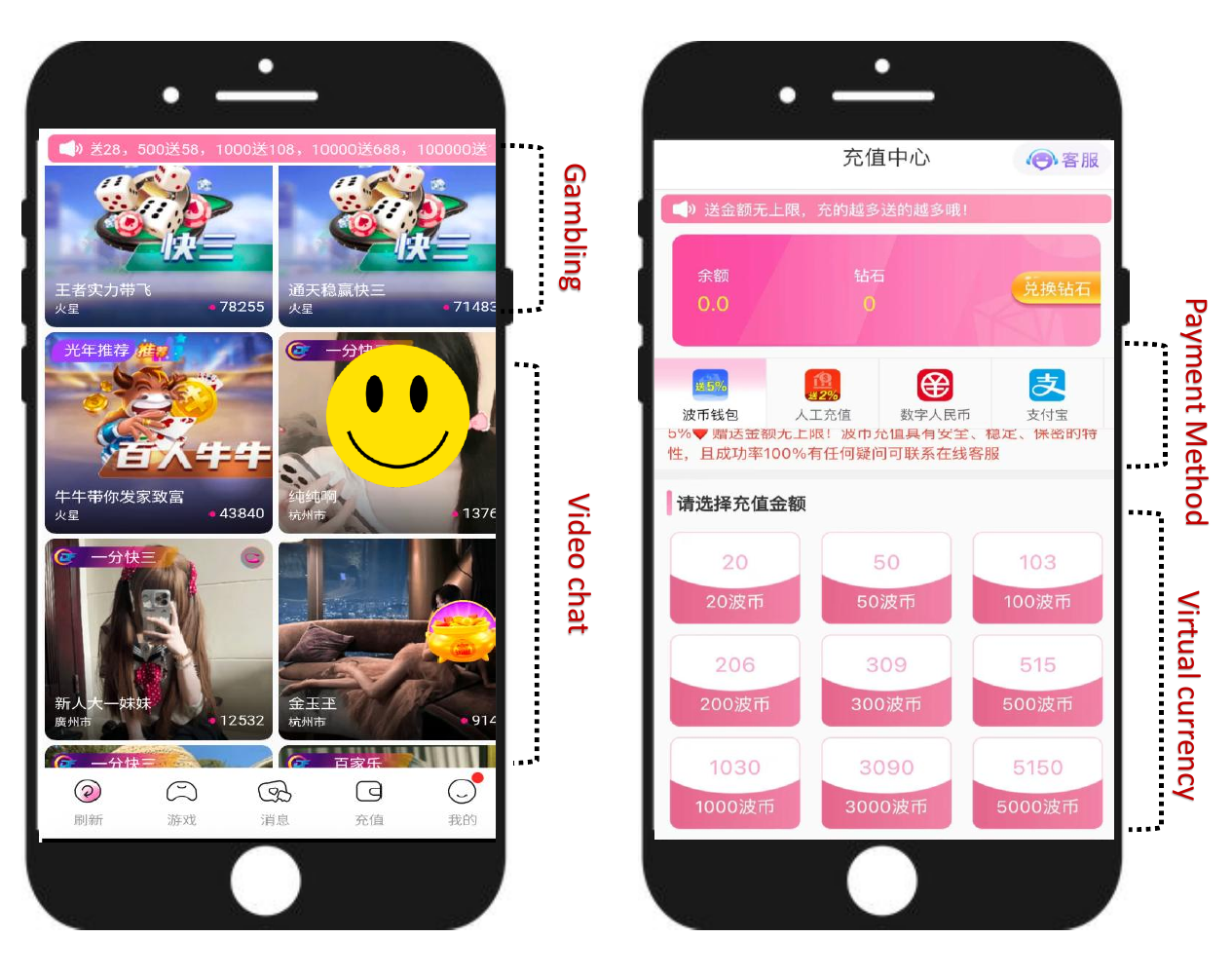}
    \caption{Example of ``\textit{Mugua Video}''}
    \label{fig:case1}
\end{figure}

\noindent\textbf{Case Study 1: Redirection to a Third-Party App.}
Our first case involves an example of being directed to a third-party app after an initial interaction. After engaging in a conversation with an individual, we were quickly asked to download an app called ``\textit{Mugua Video}'' (\textit{ehax.nzgupiwkw.mycsooqs}). As shown in \autoref{fig:case1}, at first glance, this app appeared to be a standard live-streaming platform, but upon closer inspection, it became apparent that it was part of a broader ecosystem of adult content and online gambling. The platform featured many explicit and suggestive titles of women performing live broadcasts, designed to entice users into engaging with the content. Additionally, the app was rife with gambling-related activities, which are known to be an attraction for individuals seeking quick and risky financial gains.

The app requires users to recharge their accounts in order to access the content, and during this process, users are asked to provide their phone numbers for registration. This opens the door to potential privacy violations, including the risk of personal information being sold or exploited for further malicious purposes. Moreover, the app supports multiple payment methods, including Bank and Alipay, making it easier for cybercriminals to circumvent traditional payment gateways and engage in more anonymous transactions.

\begin{table}[ht]
\centering
\scriptsize
\begin{tabular}{p{4cm}p{3.8cm}}
\toprule
\textbf{Permission} & \textbf{Purpose} \\
\midrule
\texttt{READ\_MEDIA\_VISUAL\_USER\_SELECTED} & Access visual media selected by the user \\
\rowcolor{red!10}\texttt{READ\_EXTERNAL\_STORAGE} & Read files from external storage \\
\texttt{READ\_MEDIA\_IMAGES} & Access image files on the device \\
\texttt{READ\_MEDIA\_VIDEO} & Access video files on the device \\
\texttt{READ\_MEDIA\_AUDIO} & Access audio files on the device \\
\texttt{INTERNET} & Access the internet \\
\texttt{ACCESS\_NETWORK\_STATE} & Check network connectivity \\
\rowcolor{red!10}\texttt{CAMERA} & Access the device camera \\
\rowcolor{red!10}\texttt{RECORD\_AUDIO} & Record audio through microphone \\
\rowcolor{red!10}\texttt{WRITE\_EXTERNAL\_STORAGE} & Write files to external storage \\
\texttt{FLASHLIGHT} & Control the device flashlight \\
\rowcolor{red!10}\texttt{MANAGE\_EXTERNAL\_STORAGE} & Full access to shared storage \\
\texttt{android.hardware.camera.autofocus} & Use camera autofocus (hardware feature) \\
\texttt{WAKE\_LOCK} & Prevent the device from sleeping \\
\rowcolor{red!10}\texttt{READ\_PHONE\_STATE} & Read phone state and identity \\
\rowcolor{red!10}\texttt{GET\_TASKS} & Retrieve information about running apps \\
\texttt{RECEIVE\_BOOT\_COMPLETED} & Receive system boot completed event \\
\rowcolor{red!10}\texttt{SYSTEM\_ALERT\_WINDOW} & Display over other apps \\
\rowcolor{red!10}\texttt{REQUEST\_INSTALL\_PACKAGES} & Request installing packages \\
\texttt{VIBRATE} & Control the device vibration \\
\rowcolor{red!10}\texttt{PACKAGE\_USAGE\_STATS} & Access app usage history and statistics \\
\rowcolor{red!10}\texttt{ACCESS\_COARSE\_LOCATION} & Access approximate location \\
\rowcolor{red!10}\texttt{ACCESS\_FINE\_LOCATION} & Access precise location \\
\texttt{POST\_NOTIFICATIONS} & Post notifications \\
\texttt{FOREGROUND\_SERVICE} & Run services in the foreground \\
\texttt{FOREGROUND\_SERVICE\_MEDIA\_PLAYBACK} & Play media in foreground service \\
\texttt{BLUETOOTH} & Use Bluetooth \\
\texttt{MODIFY\_AUDIO\_SETTINGS} & Change audio settings \\
\texttt{ACCESS\_WIFI\_STATE} & Check Wi-Fi state \\
\rowcolor{red!10}\texttt{CHANGE\_NETWORK\_STATE} & Change network connectivity \\
\rowcolor{red!10}\texttt{READ\_PRIVILEGED\_PHONE\_STATE} & Access restricted phone information \\
\bottomrule
\end{tabular}
\vspace{1mm}
\caption{Permissions Requested by \textit{Mugua Video}}
\label{tab:permissions}
\end{table}

A key aspect of this case is the app’s permissions. As shown in \autoref{tab:permissions}, we found that it asked for access to several potentially dangerous capabilities, such as the camera, file storage, and the ability to automatically start and install Android APK packages. While the camera permission is somewhat justifiable, considering that the app is centered around live-streaming and interactive video chats, it still presents a risk. Users engaging in such interactions may find themselves vulnerable to blackmail, especially if they are coerced into revealing intimate content or performing actions on camera that could later be used against them.

\begin{figure}
    \centering
    \includegraphics[width=1\linewidth]{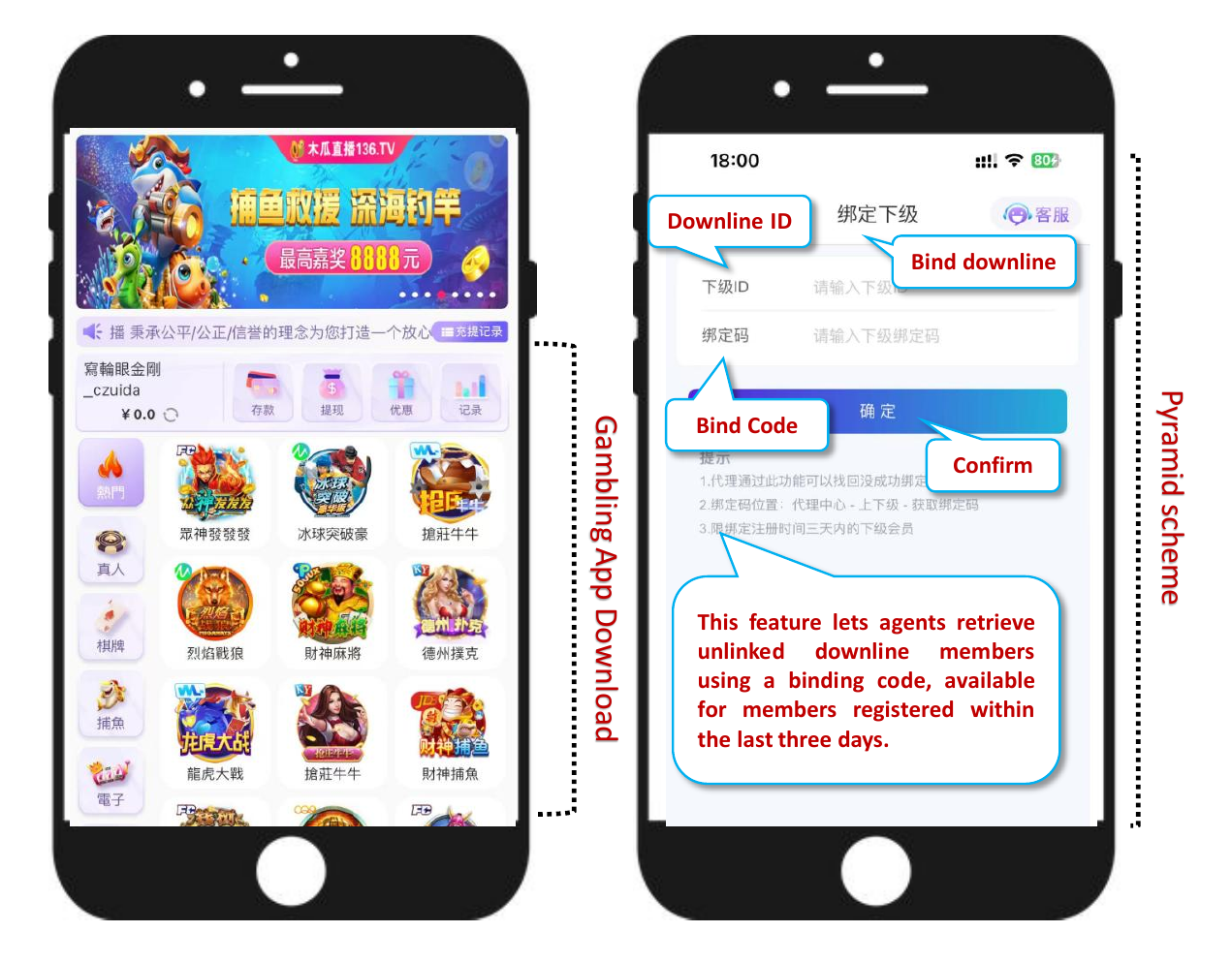}
    \caption{Gambling Software and Pyramid Scheme of ``\textit{Mugua Video}''}
    \label{fig:case2}
\end{figure}

The request for access to file storage also raises red flags, as it could allow the app to read sensitive information stored on the device. Even more concerning is the permission to install APK files automatically, which could be used to install additional software without the user’s explicit consent. This could lead to the installation of malicious apps or spyware that further compromises the device’s security.
We thoroughly reviewed the code and found no direct evidence of the app secretly installing other apps or malicious software. However, we did observe that the app subtly encourages users to install similar types of products, including other gambling software (as shown in the left part of \autoref{fig:case2}). This tactic is part of a broader strategy to create a network of interconnected apps that feed into one another, increasing user engagement and financial transactions across platforms.

Additionally, we identified a structure within the app that resembles a pyramid scheme or multi-level marketing (MLM) model. As shown in \autoref{fig:case2}, the app encourages users to invite others to download the app, offering incentives or rewards for bringing in new users. This system allows the original users to earn profits or benefits by recruiting others, creating a cycle where the more users one recruits, the more financial gain they stand to receive from lower-level participants. 
This pattern reveals a much larger ecosystem operating behind the scenes. The platform not only promotes gambling but also perpetuates a referral-based recruitment model, which can lead to exponential growth within a network. The individuals at the top of the chain benefit from the recruitment efforts of others, often at the expense of those lower down. 


\vspace{1mm}
 
\begin{figure}
    \centering
    \includegraphics[width=1\linewidth]{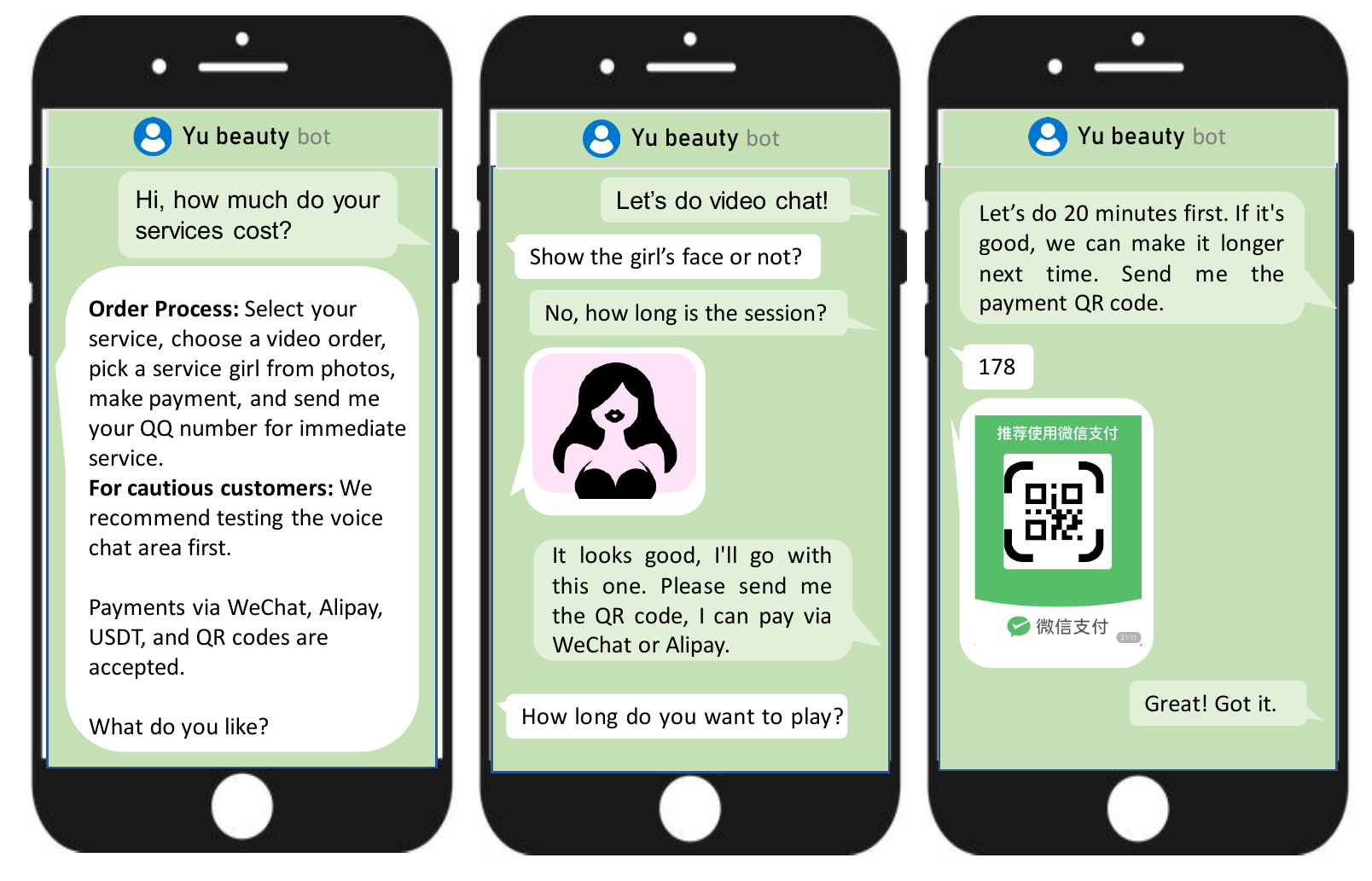}
    \caption{Interaction Process with Bot and/or Human}
    \label{fig:case3}
\end{figure}

\noindent
\textbf{Case Study 2: Interaction with a ``Bot'':}
We observed an interaction that provides interesting insights into the use of automated systems (like chatbots).  As shown in \autoref{fig:case3}, the interaction process can be summarized as follows: Upon initiating the conversation, we were immediately greeted by a bot. In the Telegram environment, a ``bot'' is typically an automated system designed to reply to messages with pre-configured responses. This bot is often used to manage customer service or assist with automated tasks. The bot's identity is explicitly indicated by the system, so we knew from the start that we were interacting with an automated system. The bot quickly introduced the service offered, outlining the process for ordering the service (choosing a video order, selecting a service girl, making a payment, and sending the QQ number for immediate service). After the initial response, the conversation quickly shifted to a more natural and human-like flow. While the initial exchange was automated, subsequent messages began to feel more personalized and seemed to be handled by a real person. 

This shift is notable because it suggests that, while the bot started the interaction, the service provider might have transferred control to a human operator during the course of the conversation. This leads us to two possible conclusions: First, the provider might be using the bot for efficiency, e.g., automating initial responses to new customers in order to handle multiple inquiries at once. Once the customer shows genuine interest, the human operator may take over to maintain the personal touch required to complete the transaction.
Second, the service provider is using a chat-based LLM (like GPT-4) to manage the conversation. If this is the case, the provider could be leveraging the advanced capabilities of LLMs to handle customer interactions without the need for a human operator. One fascinating aspect of this case is how the conversation remained highly natural and human-like, especially when the service provider sent a picture of a woman to the customer. The LLM or human operator responded with ``\textit{This looks good, I'll go with this one}.'' This is a highly personalized response that mimics real human interaction, making it harder for the target to recognize they are interacting with an automated system.

We must clearly recognize that, in the future, the integration of LLMs by service providers in chat-based scams introduces a new dimension to the traditional cat-and-mouse game between cybercriminals and defenders. LLM-based systems enable cybercriminals to automate and personalize their interactions in ways that were previously unattainable with traditional bots. These models can generate highly natural, context-aware conversations that respond to individual user inputs, making it increasingly difficult for victims to differentiate between a real human and an automated system. This enhances the likelihood of victims falling into the trap, as LLMs can adapt their responses based on the victim’s tone, behavior, or emotional cues, making the scam more personalized and convincing.
\section{Discussion}

\subsection{Board Impacts and Ultimate Objective}

 Although our work primarily focused on video sex chat scams, we believe the methodology can be extended to other forms of chat-based cybercrime, such as phishing, identity theft, and romance scams. The reason lies in the common characteristics of these scams, such as impersonation, trust-building over time, and social engineering tactics. Meanwhile, our findings provide actionable insights for designing future defense strategies. For example, Our findings suggest that LLMs can effectively engage cybercriminals (56\% success rate), but deployment should balance cost and impact. Response patterns help distinguish operator types (e.g., individuals reply quickly, platforms more slowly) enabling targeted defenses. Also, since most scams occur within five rounds, early detection is key to stopping threats before escalation. For further insights into the defense, please see \autoref{sec:defense}.


We would like to highlight that the ultimate goal of our work is to deploy bots at scale to disrupt chat-based cybercrime. By flooding cybercriminal networks with automated, context-aware interactions, we can deprive criminals of the real clients they rely on. As these bots engage with cybercriminals, without yielding actual sensitive data, cybercrimes will be forced to re-evaluate their strategies. In our case, this will at least cause service providers to waste time chatting with 30 virtual individuals. The cascading effect will create a disruptive force that makes cybercrimes less profitable, forcing them to shut down or adopt entirely new methods. With sufficient scale, these bots can be an essential part of countering chat-based cybercrime at its core.

\vspace{1mm}
\noindent\textbf{Ethics Concerns:}
To address the ethical concerns associated with our experiment, we took several steps to ensure the research complied with ethical guidelines, legal standards, and protected the privacy of all parties involved. Here's how we tackled these concerns: \looseness=-1
\begin{itemize}[partopsep=2pt, topsep=-\parskip, parsep=2pt, itemsep=2pt, leftmargin=*]
    \item \textbf{IRB Approval and Legal Consultation}: Our experiment underwent comprehensive review and approval by the IRB, ensuring it adhered to ethical guidelines for research involving human subjects. Because we do not collect personally identifiable information (PII), we do need to formally notify participants about their involvement in the work. Additionally, to avoid potential legal risks and ethical issues, we worked closely with a professor from our university’s law school. His ongoing guidance ensured that we followed legal protocols throughout the experiment, particularly regarding privacy, consent, and the use of sensitive data.
    \item \textbf{Data Management}: During the data collection process, we anonymized all collected data to ensure that no personally identifiable information of service providers was retained. This step was essential to protect their privacy. After completing the experiment, we deleted all the collected data to ensure that no sensitive information was retained. Throughout the experiment, we ensured that no information (even the data is non-PII) was shared with any third parties, minimizing the risk of privacy breaches.
    \item \textbf{Handling Sensitive Content and Photos}: Some service providers may send photos during interactions, which could include sensitive content, such as images with faces or/and private body parts. To address this, we did not save any of these photos. Moreover, we considered the possibility that some of these images might be sourced from the internet and not involve real individuals (e.g., AI generated). In consultation with legal experts from our law school, we were advised that, for service providers in the adult service sector,  they often share such content in public spaces, like social platforms or group chats, and in their professional context, they may not consider it to be private. Therefore, we ensured our actions were in line with legal interpretations, treating the shared content with appropriate caution.
    \item \textbf{Preventing Harmful Content from LLM:} To prevent the LLM from generating harmful content, such as insulting or offensive messages to the service providers, we implemented a manual review process. Each message generated by the LLM was carefully reviewed by our team before being sent. This step ensured that the content did not harm or insult anyone involved in the experiment. We limited the number of conversations to just over 50 to ensure thorough monitoring and review of each interaction.
     \item \textbf{Non-participation in the Service}: We took every precaution to avoid participating in any illegal or unethical activities. Specifically, we did not purchase any of the services offered by the service providers nor did we engage in any explicit activities, such as engaging in video sex chats. The purpose of our experiment was purely to analyze and understand the dynamics of chat-based cybercrime, not to participate in or encourage any illicit activities.
\end{itemize}

\section{Related Work}

LLM have emerged as potent tools for combating various forms of cybercrime, extending beyond general cybersecurity applications to address specific criminal activities in the digital realm \cite{anonymous2024comprehensive}. For example, 
In the domain of fraud detection, LLMs demonstrate significant efficacy in analyzing linguistic patterns and contextual cues that signal fraudulent intent. Recent evaluations of GPT-based models reveal their ability to identify elements of phishing and scams, though continued refinement is necessary to address rapidly evolving criminal tactics \cite{anonymous2024detecting,korkanti2024enhancing,yang2025fraud,zhao2024explain,shen2025warned,dahiphale2024enhancing,amincommerce,ma2025teleantifraud,shen2025combating,chakraborty2024detoxbench}, achieving high accuracy in detecting fraudulent communications and impersonation attempts in real-time settings \cite{anonymous2025advanced}.
For phishing and social engineering countermeasures, LLMs excel at identifying manipulative linguistic techniques, including emotional appeals, false urgency, and impersonation tactics commonly employed in these attacks~\cite{anonymous2024defending,greco2024david, koide2024chatspamdetector, hafzullah2024llm, afane2024next, fairbanks2024generating, heiding2024devising, garde2024security, shyni2025generative, chataut2024can, trad2024prompt, ahmed2025comprehensive, li2024knowphish, desolda2024apollo, roy2024chatbots}. The DetoxBench framework provides a comprehensive evaluation methodology for assessing LLM performance in detecting abusive and fraudulent language across diverse scenarios, highlighting both strengths and limitations in current implementations~\cite{anonymous2024detoxbench}.
LLMs also contribute to ransomware and malware defense through code analysis. Code-specialized models like CodeLlama demonstrate particular efficacy in detecting security flaws and suggesting remediation strategies \cite{anonymous2024when,zhou2025srdc, jelodar2025large, altamimi2024detecting, oh2024volgpt, kumar2024prompt, hossain2024malicious, omar2024harnessing, patsakis2024assessing, wang2023using, yu2024maltracker, khanllms}. In the realm of identity theft prevention, LLMs analyze behavioral patterns in user interactions to flag anomalous activities that may indicate compromised credentials or account takeovers \cite{anonymous2024comprehensive}.

These approaches, while effective in isolated environments, typically treat the LLM as a passive classifier, an intelligent filter applied post hoc to existing artifacts. In contrast, our work represents a shift from passive analysis to active engagement. We are the first to operationalize LLMs as real-time participants in adversarial communication ecosystems, specifically targeting the chat-based cybercrime landscape.


\section{Conclusion}

In this work, we present LURE, a system that leverages Large Language Models as active agents within adversarial chat environments. Unlike traditional passive detection approaches, LURE engages cybercriminals in real time, mimicking human behavior to extract valuable intelligence. Through a large-scale deployment targeting illicit video chat scams on Telegram, we show that LLMs can sustain realistic conversations, often without being detected as bots. Our findings reveal consistent behavioral patterns in scam operations and demonstrate the feasibility of using LLMs to infiltrate and disrupt trust-based cybercrime ecosystems. This approach opens a new direction for proactive, scalable, and AI-driven defense strategies.

\bibliographystyle{plain}
\bibliography{ref, ref_liyue}

\begin{thebibliography}{10}

\bibitem{afane2024next}
Khalifa Afane, Wenqi Wei, Ying Mao, Junaid Farooq, and Juntao Chen.
\newblock Next-generation phishing: How llm agents empower cyber attackers.
\newblock In {\em 2024 IEEE International Conference on Big Data (BigData)}, pages 2558--2567. IEEE, 2024.

\bibitem{ahmed2025comprehensive}
Sajjad Ahmed, Irshad~Ahmed Sumra, Ijaz Khan, and Hadi Abdullah.
\newblock A comprehensive review on the role of ai in phishing detection mechanisms.
\newblock {\em Journal of Computing \& Biomedical Informatics}, 8(02), 2025.

\bibitem{altamimi2024detecting}
Shahad Altamimi and Mohammad Ababneh.
\newblock Detecting spam and malware using bert and llms.
\newblock In {\em 2024 25th International Arab Conference on Information Technology (ACIT)}, pages 1--6. IEEE, 2024.

\bibitem{amincommerce}
Usuf Amin, Md~Abu Sayed, and Nishat Anjum.
\newblock E-commerce security: Leveraging large language models for fraud detection and data protection.

\bibitem{anonymous2024comprehensive}
Anonymous.
\newblock A comprehensive overview of large language models (llms) for cyber defences: Opportunities and directions.
\newblock {\em arXiv preprint arXiv:2405.14487}, 2024.

\bibitem{anonymous2024defending}
Anonymous.
\newblock Defending against social engineering attacks in the age of llms.
\newblock {\em arXiv preprint arXiv:2406.12263}, 2024.

\bibitem{anonymous2024detecting}
Anonymous.
\newblock Detecting scams using large language models.
\newblock {\em arXiv preprint arXiv:2402.03147}, 2024.

\bibitem{anonymous2024detoxbench}
Anonymous.
\newblock Detoxbench: Benchmarking large language models for multitask fraud \& abuse detection.
\newblock {\em arXiv preprint arXiv:2409.06072}, 2024.

\bibitem{anonymous2024when}
Anonymous.
\newblock When llms meet cybersecurity: A systematic literature review.
\newblock {\em arXiv preprint arXiv:2405.03644}, 2024.

\bibitem{anonymous2025advanced}
Anonymous.
\newblock Advanced real-time fraud detection using rag-based llms.
\newblock {\em arXiv preprint arXiv:2501.15290}, 2025.

\bibitem{bilz2023tainted}
Alexander Bilz, Lynsay~A Shepherd, and Graham~I Johnson.
\newblock Tainted love: A systematic literature review of online romance scam research.
\newblock {\em Interacting with Computers}, 35(6):773--788, 2023.

\bibitem{chakraborty2024detoxbench}
Joymallya Chakraborty, Wei Xia, Anirban Majumder, Dan Ma, Walid Chaabene, and Naveed Janvekar.
\newblock Detoxbench: Benchmarking large language models for multitask fraud \& abuse detection.
\newblock {\em arXiv preprint arXiv:2409.06072}, 2024.

\bibitem{chataut2024can}
Robin Chataut, Prashnna~Kumar Gyawali, and Yusuf Usman.
\newblock Can ai keep you safe? a study of large language models for phishing detection.
\newblock In {\em 2024 IEEE 14th Annual Computing and Communication Workshop and Conference (CCWC)}, pages 0548--0554. IEEE, 2024.

\bibitem{dahiphale2024enhancing}
Devendra Dahiphale, Naveen Madiraju, Justin Lin, Rutvik Karve, Monu Agrawal, Anant Modwal, Ramanan Balakrishnan, Shanay Shah, Govind Kaushal, Priya Mandawat, et~al.
\newblock Enhancing trust and safety in digital payments: An llm-powered approach.
\newblock In {\em 2024 IEEE International Conference on Big Data (BigData)}, pages 4854--4863. IEEE, 2024.

\bibitem{desolda2024apollo}
Giuseppe Desolda, Francesco Greco, and Luca Vigan{\`o}.
\newblock Apollo: A gpt-based tool to detect phishing emails and generate explanations that warn users.
\newblock {\em arXiv preprint arXiv:2410.07997}, 2024.

\bibitem{fairbanks2024generating}
Jeffrey Fairbanks and Edoardo Serra.
\newblock Generating phishing attacks and novel detection algorithms in the era of large language models.
\newblock In {\em 2024 IEEE International Conference on Big Data (BigData)}, pages 2314--2319. IEEE, 2024.

\bibitem{fbi2023sextortion}
{Federal Bureau of Investigation}.
\newblock Sextortion: A growing threat targeting minors.
\newblock \url{https://www.fbi.gov/contact-us/field-offices/nashville/news/sextortion-a-growing-threat-targeting-minors}, 2022.
\newblock Accessed: 2025-04-15.

\bibitem{garde2024security}
Tanishq Garde, Manas Rathi, Sanskar Dubey, and SS~Narkhede.
\newblock Security posture detection using llm.
\newblock In {\em AIP Conference Proceedings}, volume 3156. AIP Publishing, 2024.

\bibitem{greco2024david}
Francesco Greco, Giuseppe Desolda, Andrea Esposito, Alessandro Carelli, et~al.
\newblock David versus goliath: Can machine learning detect llm-generated text? a case study in the detection of phishing emails.
\newblock In {\em The Italian Conference on CyberSecurity}, 2024.

\bibitem{hafzullah2024llm}
{\.I}{\c{S}}~Hafzullah.
\newblock Llm-driven sat impact on phishing defense: A cross-sectional analysis.
\newblock In {\em 2024 12th International Symposium on Digital Forensics and Security (ISDFS)}, pages 1--5. IEEE, 2024.

\bibitem{heiding2024devising}
Fredrik Heiding, Bruce Schneier, Arun Vishwanath, Jeremy Bernstein, and Peter~S Park.
\newblock Devising and detecting phishing emails using large language models.
\newblock {\em IEEE Access}, 2024.

\bibitem{hodges2014alan}
Andrew Hodges.
\newblock {\em Alan Turing: The Enigma: The Book That Inspired the Film" The Imitation Game"}.
\newblock Princeton University Press, 2014.

\bibitem{hossain2024malicious}
Al~Amin Hossain, Mithun~Kumar PK, Junjie Zhang, and Fathi Amsaad.
\newblock Malicious code detection using llm.
\newblock In {\em NAECON 2024-IEEE National Aerospace and Electronics Conference}, pages 414--416. IEEE, 2024.

\bibitem{jelodar2025large}
Hamed Jelodar, Samita Bai, Parisa Hamedi, Hesamodin Mohammadian, Roozbeh Razavi-Far, and Ali Ghorbani.
\newblock Large language model (llm) for software security: Code analysis, malware analysis, reverse engineering.
\newblock {\em arXiv preprint arXiv:2504.07137}, 2025.

\bibitem{khanllms}
Muhammad Al-Zafar Khan, Jamal Al-Karaki, and Marwan Omar.
\newblock Llms for malware detection: Review, framework design, and countermeasure approaches.
\newblock {\em Framework Design, and Countermeasure Approaches}.

\bibitem{koide2024chatspamdetector}
Takashi Koide, Naoki Fukushi, Hiroki Nakano, and Daiki Chiba.
\newblock Chatspamdetector: Leveraging large language models for effective phishing email detection.
\newblock {\em arXiv preprint arXiv:2402.18093}, 2024.

\bibitem{korkanti2024enhancing}
Sukanth Korkanti.
\newblock Enhancing financial fraud detection using llms and advanced data analytics.
\newblock In {\em 2024 2nd International Conference on Self Sustainable Artificial Intelligence Systems (ICSSAS)}, pages 1328--1334. IEEE, 2024.

\bibitem{kumar2024prompt}
Neha~Mohan Kumar, Fahmida~Tasnim Lisa, and Sheikh~Rabiul Islam.
\newblock Prompt chaining-assisted malware detection: A hybrid approach utilizing fine-tuned llms and domain knowledge-enriched cybersecurity knowledge graphs.
\newblock In {\em 2024 IEEE International Conference on Big Data (BigData)}, pages 1672--1677. IEEE, 2024.

\bibitem{li2024knowphish}
Yuexin Li, Chengyu Huang, Shumin Deng, Mei~Lin Lock, Tri Cao, Nay Oo, Hoon~Wei Lim, and Bryan Hooi.
\newblock $\{$KnowPhish$\}$: Large language models meet multimodal knowledge graphs for enhancing $\{$Reference-Based$\}$ phishing detection.
\newblock In {\em 33rd USENIX Security Symposium (USENIX Security 24)}, pages 793--810, 2024.

\bibitem{ma2025teleantifraud}
Zhiming Ma, Peidong Wang, Minhua Huang, Jingpeng Wang, Kai Wu, Xiangzhao Lv, Yachun Pang, Yin Yang, Wenjie Tang, and Yuchen Kang.
\newblock Teleantifraud-28k: A audio-text slow-thinking dataset for telecom fraud detection.
\newblock {\em arXiv preprint arXiv:2503.24115}, 2025.

\bibitem{nguyen2017exploring}
Trong~Duc Nguyen, Anh~Tuan Nguyen, Hung~Dang Phan, and Tien~N Nguyen.
\newblock Exploring api embedding for api usages and applications.
\newblock In {\em 2017 IEEE/ACM 39th International Conference on Software Engineering (ICSE)}, pages 438--449. IEEE, 2017.

\bibitem{oh2024volgpt}
Dong~Bin Oh, Donghyun Kim, and Huy~Kang Kim.
\newblock volgpt: Evaluation on triaging ransomware process in memory forensics with large language model.
\newblock {\em Forensic Science International: Digital Investigation}, 49:301756, 2024.

\bibitem{omar2024harnessing}
Marwan Omar, Hewa~Majeed Zangana, Jamal~N Al-Karaki, and Derek Mohammed.
\newblock Harnessing llms for iot malware detection: A comparative analysis of bert and gpt-2.
\newblock In {\em 2024 8th International Symposium on Multidisciplinary Studies and Innovative Technologies (ISMSIT)}, pages 1--6. IEEE, 2024.

\bibitem{patsakis2024assessing}
Constantinos Patsakis, Fran Casino, and Nikolaos Lykousas.
\newblock Assessing llms in malicious code deobfuscation of real-world malware campaigns.
\newblock {\em Expert Systems with Applications}, 256:124912, 2024.

\bibitem{roy2024chatbots}
Sayak~Saha Roy, Poojitha Thota, Krishna~Vamsi Naragam, and Shirin Nilizadeh.
\newblock From chatbots to phishbots?: Phishing scam generation in commercial large language models.
\newblock In {\em 2024 IEEE Symposium on Security and Privacy (SP)}, pages 36--54. IEEE, 2024.

\bibitem{shah2021cybercrime}
Balkrishna Shah, Nitu Sharma, Saloni Bandgar, and Sainath Patil.
\newblock Cybercrime prevention on social media.
\newblock {\em International Journal of Engineering Research \& Technology (IJERT)}, 10(03), 2021.

\bibitem{shen2025combating}
Zitong Shen, Kangzhong Wang, Youqian Zhang, Grace Ngai, and Eugene~Yujun Fu.
\newblock Combating phone scams with llm-based detection: Where do we stand?(student abstract).
\newblock In {\em Proceedings of the AAAI Conference on Artificial Intelligence}, volume~39, pages 29487--29489, 2025.

\bibitem{shen2025warned}
Zitong Shen, Sineng Yan, Youqian Zhang, Xiapu Luo, Grace Ngai, and Eugene~Yujun Fu.
\newblock " it warned me just at the right moment": Exploring llm-based real-time detection of phone scams.
\newblock {\em arXiv preprint arXiv:2502.03964}, 2025.

\bibitem{shyni2025generative}
C~Emilin Shyni et~al.
\newblock Generative ai-based phishing text generation using hybrid prompt design with heuristic algorithm for multimodal phishing detection.
\newblock {\em International Journal of Intelligent Engineering \& Systems}, 18(2), 2025.

\bibitem{singh2020intelligent}
Amanpreet Singh and Maninder Kaur.
\newblock Intelligent content-based cybercrime detection in online social networks using cuckoo search metaheuristic approach.
\newblock {\em The Journal of Supercomputing}, 76(7):5402--5424, 2020.

\bibitem{soomro2019social}
Tariq~Rahim Soomro and Mumtaz Hussain.
\newblock Social media-related cybercrimes and techniques for their prevention.
\newblock {\em Appl. Comput. Syst.}, 24(1):9--17, 2019.

\bibitem{telegram}
{Telegram Messenger LLP}.
\newblock Telegram — fast and secure messaging app.
\newblock \texttt{https://telegram.org/}, 2024.
\newblock \url{https://telegram.org/}.

\bibitem{wechat}
{Tencent}.
\newblock Wechat — connecting a billion people.
\newblock \texttt{https://www.wechat.com/}, 2024.
\newblock \url{https://www.wechat.com/}.

\bibitem{trad2024prompt}
Fouad Trad and Ali Chehab.
\newblock Prompt engineering or fine-tuning? a case study on phishing detection with large language models.
\newblock {\em Machine Learning and Knowledge Extraction}, 6(1):367--384, 2024.

\bibitem{wang2023using}
Fang Wang.
\newblock Using large language models to mitigate ransomware threats, 2023.

\bibitem{whatsapp}
{WhatsApp LLC}.
\newblock Whatsapp — simple. secure. reliable messaging.
\newblock \texttt{https://www.whatsapp.com/}, 2024.
\newblock \url{https://www.whatsapp.com/}.

\bibitem{yang2025fraud}
Shu Yang, Shenzhe Zhu, Zeyu Wu, Keyu Wang, Junchi Yao, Junchao Wu, Lijie Hu, Mengdi Li, Derek~F Wong, and Di~Wang.
\newblock Fraud-r1: A multi-round benchmark for assessing the robustness of llm against augmented fraud and phishing inducements.
\newblock {\em arXiv preprint arXiv:2502.12904}, 2025.

\bibitem{yu2024maltracker}
Zeliang Yu, Ming Wen, Xiaochen Guo, and Hai Jin.
\newblock Maltracker: A fine-grained npm malware tracker copiloted by llm-enhanced dataset.
\newblock In {\em Proceedings of the 33rd ACM SIGSOFT International Symposium on Software Testing and Analysis}, pages 1759--1771, 2024.

\bibitem{zhao2024explain}
Yong Zhao, Can Liu, Shubham Agrawal, and Chiranjeet Chetia.
\newblock Explain fraud detection method and system via fraud graph and features similarity based knn few-shot example-prompts.
\newblock 2024.

\bibitem{zhou2025srdc}
Ce~Zhou, Yilun Liu, Weibin Meng, Shimin Tao, Weinan Tian, Feiyu Yao, Xiaochun Li, Tao Han, Boxing Chen, and Hao Yang.
\newblock Srdc: Semantics-based ransomware detection and classification with llm-assisted pre-training.
\newblock In {\em Proceedings of the AAAI Conference on Artificial Intelligence}, volume~39, pages 28566--28574, 2025.

\end{thebibliography}
\appendix

\section{Defense Strategy}
\label{sec:defense}
 
These lessons learned provide strong foundations for designing future defense strategies.  To be more specific: 

\begin{itemize}[partopsep=2pt, topsep=-\parskip, parsep=2pt, itemsep=2pt, leftmargin=*]
    \item  \textbf{LLM Interaction Success vs. Deployment Strategy.}  Our experiments show that LLMs have a success rate of at least 56\% in interacting with cybercriminals. This demonstrates their effectiveness in simulating victims. However, a trade-off exists between economic losses caused by cybercrime and the prevention achievable through LLM deployment. By calculating this, we can design more efficient deployment strategies to maximize the effectiveness of LLMs in mitigating financial losses from scams.
   \item \textbf{Response Time Differences in Individual vs. Platform Operators.} We observed differences in response times and interaction patterns between individual and platform operators. Individual operators respond quickly, while platform operators have slower but more consistent responses. This insight allows for targeted defense strategies: fast responses may indicate individual operators, requiring frequent monitoring, while platforms require long-term defense strategies due to their consistent engagement.

\item \textbf{Initial Response Time, Interaction Rounds, and Success Rate.} Our experiment shows that most criminals respond quickly, and almost all scams are completed within five rounds. This suggests that cybercrimes rely on fast engagement. Defense systems should focus on the early stages of interaction, using computational resources to identify and stop high-risk conversations before they escalate, conserving resources while preventing crime from progressing.

\end{itemize}

\end{document}